\documentclass[aps,pra,reprint,showpacs]{revtex4-1}
\usepackage{amsmath}
\usepackage{SIunits}
\usepackage{graphicx}% Include figure files

\begin{document}
%
%
% Define new commands:
%
\newcommand{\alphaR}[0]{\ensuremath{\alpha_{\mathrm{R}}}}
\newcommand{\alphaI}[0]{\ensuremath{\alpha_{\mathrm{I}}}}
\renewcommand{\a}[0]{\ensuremath{\hat{a}}}
\newcommand{\adag}[0]{\ensuremath{\hat{a}^{\dagger}}}
\newcommand{\betaI}[0]{\ensuremath{\beta_\mathrm{I}}}
\newcommand{\betaR}[0]{\ensuremath{\beta_\mathrm{R}}}
\renewcommand{\c}[0]{\ensuremath{\hat{c}}}
\newcommand{\cdag}[0]{\ensuremath{\hat{c}^{\dagger}}}
\newcommand{\CovMat}[0]{\ensuremath{\boldsymbol\gamma}}
\newcommand{\dens}[0]{\ensuremath{\hat{\rho}}}
\newcommand{\densin}[0]{\ensuremath{\hat{\rho}_{\mathrm{in}}}}
\newcommand{\densout}[0]{\ensuremath{\hat{\rho}_{\mathrm{out}}}}
\newcommand{\densSTS}[0]{\ensuremath{\hat{\rho}_{\mathrm{STS}}}}
\newcommand{\densRSTS}[0]{\ensuremath{\hat{\rho}_{\mathrm{RSTS}}}}
\newcommand{\Fq}[0]{\ensuremath{F_q}}
\renewcommand{\H}[0]{\ensuremath{\hat{H}}}
\newcommand{\Hext}[0]{\ensuremath{\hat{H}_{\mathrm{ext}}}}
\newcommand{\Hint}[0]{\ensuremath{\hat{H}_{\mathrm{I}}}}
\renewcommand{\Im}[0]{\ensuremath{\mathrm{Im}}}
\newcommand{\mat}[1]{\ensuremath{\mathbf{#1}}}
\newcommand{\mean}[1]{\ensuremath{\langle#1\rangle}}
\newcommand{\pauli}[0]{\ensuremath{\hat{\sigma}}}
\newcommand{\Pa}[0]{\ensuremath{\hat{P}_{\mathrm{c}}}}
\newcommand{\Pmax}[0]{\ensuremath{P_{\mathrm{max}}}}
\renewcommand{\Re}[0]{\ensuremath{\mathrm{Re}}}
\renewcommand{\S}[0]{\ensuremath{\hat{S}}}
\newcommand{\Var}[0]{\ensuremath{\mathrm{Var}}}
\newcommand{\Covar}[0]{\ensuremath{\mathrm{Cov}}}
\renewcommand{\vec}[1]{\ensuremath{\mathbf{#1}}}
\newcommand{\Xin}[0]{\ensuremath{\hat{X}_{\mathrm{in}}}}
\newcommand{\Xout}[0]{\ensuremath{\hat{X}_{\mathrm{out}}}}
\newcommand{\Pin}[0]{\ensuremath{\hat{P}_{\mathrm{in}}}}
\newcommand{\Pout}[0]{\ensuremath{\hat{P}_{\mathrm{out}}}}
\newcommand{\bin}[0]{\ensuremath{\hat{b}_{\mathrm{in}}}}
\newcommand{\ket}[1]{\ensuremath{|#1\rangle}}
\newcommand{\bra}[1]{\ensuremath{\langle#1|}}

\title{Fidelity of Fock-state-encoded qubits subjected to continuous
  variable Gaussian processes}

\author{Brian Julsgaard}
\email{brianj@phys.au.dk}

\author{Klaus M{\o}lmer}
\affiliation{Department of Physics and Astronomy, Aarhus University, Ny
  Munkegade 120, DK-8000 Aarhus C, Denmark.}

%Collaboration name if desired (requires use of superscriptaddress
%option in \documentclass). \noaffiliation is required (may also be
%used with the \author command).
%\collaboration can be followed by \email, \homepage, \thanks as well.
%\collaboration{}
%\noaffiliation

\date{\today}

\begin{abstract}
  When a harmonic oscillator is under the influence of a Gaussian
  process such as linear damping, parametric gain, and linear coupling
  to a thermal environment, its coherent states are transformed into
  states with Gaussian Wigner function. Qubit states can be encoded in
  the $\ket{0}$ and $\ket{1}$ Fock states of a quantum harmonic
  oscillator, and it is relevant to know the fidelity of the output
  qubit state after a Gaussian process on the oscillator. In this
  paper we present a general expression for the average qubit fidelity
  in terms of the first and second moments of the output from input
  coherent states subjected to Gaussian processes.
\end{abstract}

\pacs{03.67.-a, 03.67.Hk}

\maketitle

\section{Introduction}
\label{sec:Introduction}
In analogy with the classical bit in computer science, the qubit forms
the most basic building block within the field of quantum information
\cite{DiVincenzo.Science.270.255(1995)}. In order to perform quantum
computation one must, among other tasks, be able to initialize,
manipulate, and read out the information encoded in qubits, and in a
scalable implementation it is necessary to store quantum information
and transport it from one place or medium to another. Such
operations are applied in quantum memories for few-photon light pulses in single
atoms \cite{Boozer.PhysRevLett.98.193601(2007),
  Specht.Nature.473.190(2011)} and in quantum teleportation between
similar qubits \cite{Riebe.Nature.429.734(2004),
  Barrett.Nature.429.737(2004)}.  In parallel to qubit-based quantum
information science there has also been attention to
continuous-variable versions of quantum computation
\cite{Braunstein.RevModPhys.77.513(2005)}, quantum teleportation
\cite{Furusawa.Science.282.706(1998), Sherson.Nature.443.557(2006),
  Krauter.NaturePhys.9.400(2013)}, and quantum memories
\cite{Julsgaard.Nature.432.482(2004),
  Honda.PhysRevLett.100.093601(2008),
  Appel.PhysRevLett.100.093602(2008),
  deRiedmatten.Nature.456.773(2008),
  Hedges.Nature.465.1052(2010)}. These protocols can be implemented
in, e.g., quadrature variables of electromagnetic fields
\cite{Furusawa.Science.282.706(1998)}, atomic or solid state ensembles
of spins \cite{Hammerer.RevModPhys.82.1041(2010),
  Tittel.LaserPhotonRev.4.244(2010)}, or vibration modes of
nano-mechanical oscillators \cite{Kleckner.Nature.444.75(2006),
  Thompson.Nature.452.72(2008),
  Kippenberg.PhysRevLett.95.033901(2005),
  Regal.NaturePhys.4.555(2008), Hunger.PhysRevLett.104.143002(2010)},
which are all exact or excellent approximate realizations of the quantum
harmonic oscillator.

While the discrete and continuous variable versions of quantum
information originally seemed as detached scientific domains, there
have been demonstrations of single light quanta, discrete in nature, transferred
into the collective spin degrees of freedom of macroscopic atomic
ensembles, which are continuous in nature
\cite{Chou.Nature.438.828(2005), Chaneliere.Nature.438.833(2005),
  Eisaman.Nature.438.837(2005)}. More recently, quantum memories for
photonic qubits have been implemented benefiting from the increased
collective interaction strength of atomic ensembles compared to single
atoms \cite{Kubo.PhysRevLett.107.220501(2011),
  Clausen.PhysRevLett.108.190503(2012),
  Gundogan.PhysRevLett.108.190504(2012)}. In connection with the use
of hybrid physical systems for quantum information processing, multiple
proposals exist, making use of the interconnection of mesoscopic qubit
degrees of freedom and the continuous variables of ensembles of
microscopic systems, nano-mechanical resonators and quantized field
modes \cite{Xiang.RevModPhys.85.623(2013),
  Rabl.PhysRevLett.97.033003(2006),
  Petrosyan.PhysRevA.79.040304R(2009),
  Wesenberg.PhysRevLett.103.070502(2009),
  Marcos.PhysRevLett.105.210501(2010),
  Kubo.PhysRevLett.107.220501(2011), OConnell.Nature.464.697(2010),
  Arcizet.NaturePhys.7.879(2011), Pirkkalainen.Nature.494.211(2013)}.

The present manuscript addresses an important question in this
context: If the transformation properties of continuous variables are
known for a particular process in a given physical system, then what
can be said about a qubit encoded into the same system and subjected
to the same transformation? Specifically, if a harmonic oscillator is
subjected to a Gaussian process, characterized by its effect on the
first and second moments of the conjugate variables $\hat{X}$ and
$\hat{P}$, we present a general formula for the qubit fidelity,
i.e.~the probability that the input state of a qubit encoded into the
$\ket{0}$ and $\ket{1}$ Fock states of the harmonic oscillator
coincide with the output state after the system has been exposed to
the process. A Gaussian process can be characterized completely by its
action on a small set of coherent states
\cite{Wang.PhysRevA.88.022101(2013)}, and as pointed out in
Ref.~\cite{Sherson.Nature.443.557(2006)} this is easier than preparing
qubit states for experimental determination of its fidelity. Also, from a
theoretical perspective, as exemplified by
Ref.~\cite{Julsgaard.PhysRevLett.110.250503(2013)}, the quantum memory
fidelity for qubit states can be calculated more easily in a
multi-mode set-up by using coherent input states and accounting solely
for the first and second moments of the physical variables involved.

The paper is arranged as follows: In Sec.~\ref{sec:LinearMap} we show
how the observed first and second moments of output states, following
from application of coherent input states to the process, yield a
convenient parametrization of the Gaussian process. In
Sec.~\ref{sec:QubitFidelity} we derive the average fidelity over all
qubit states encoded in the $|0\rangle$ and $|1\rangle$ Fock states
and subjected to the Gaussian process.  In
Sec.~\ref{sec:Specific_examples}, we present some specific examples,
and in Sec.~\ref{sec:summary} we conclude the manuscript.

\section{Parametrizing the Gaussian  process}
\label{sec:LinearMap}
\begin{figure}[t]
  \centering
  \includegraphics{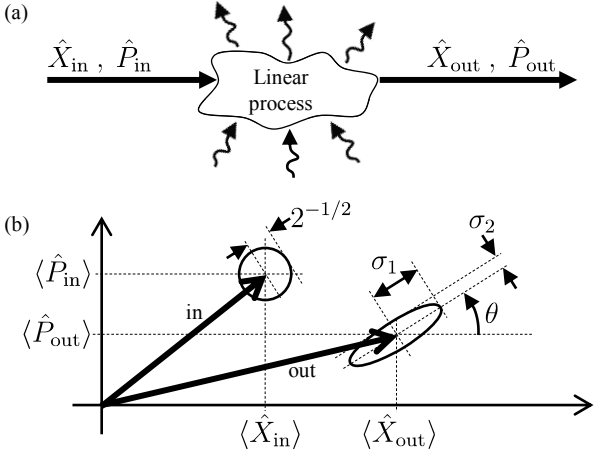}
  \caption{(a) A single mode of a harmonic oscillator is subjected to
    a Gaussian process, which maps the input observables $\Xin$ and
    $\Pin$ into the output variables $\Xout$ and $\Pout$, of the same
    or a different quantum system.  The wavy arrows represent, e.g.,
    absorption losses or addition of thermal noise associated with the
    possible coupling to environment degrees of freedom. (b) Schematic
    view of the transformation of a coherent input state with
    $\Var(\Xin) = \Var(\Pin) = \frac{1}{2}$. The solid arrows indicate
    the mean values, and the circle and the ellipse show the standard
    deviation of the continuous quadrature variables in the
    $xp$-plane. For the output state, $\theta$ denotes the angle
    between the $x$-axis and the major axis of the uncertainty ellipse
    with $\sigma_1 \ge \sigma_2$.}
  \label{fig:BasicSetup}
\end{figure}
We consider a process, which maps an input quantum state of a single
harmonic oscillator to an output state on the same or a different
oscillator. For instance this could represent the storage of a
radiation-field state into polarization modes of a spin ensemble
\cite{Julsgaard.Nature.432.482(2004)}, or it could represent the
teleportation of the quadrature amplitudes from one laser beam to
another
\cite{Furusawa.Science.282.706(1998)}. Figure~\ref{fig:BasicSetup}(a)
shows schematically how this process transforms the input harmonic
oscillator mode $(\Xin,\Pin)$ into the output mode $(\Xout,\Pout)$
under the possible influence of the environment. For any input state
density matrix $\densin$ this process is mathematically described by a
map, $\densout = E(\densin)$, and our task is to (i) establish a
suitable parametrization of this map, and to (ii) calculate the
fidelity when a qubit state is subjected to the process.

The restriction to Gaussian processes relies on two assumptions about
the map $E$. The first one is that it is linear in the sense that our
input and output harmonic oscillator modes couple linearly to each
other and to all auxiliary reservoir modes. Thus we assume that
$(\Xin,\Pin)$, $(\Xout,\Pout)$, and the reservoir variables
$(\hat{x}_j,\hat{p}_j)$ with $j=1,\ldots,n$, obey the equation:
\begin{equation}
\label{eq:General_linear_transform}
  \begin{bmatrix}
    \Xout \\ \Pout \\ \hat{x}_1^{\mathrm{out}} \\ \hat{p}_1^{\mathrm{out}} \\
    \vdots \\ \hat{x}_n^{\mathrm{out}} \\ \hat{p}_n^{\mathrm{out}}
  \end{bmatrix}
  = \mat{A}
  \begin{bmatrix}
    \Xin \\ \Pin \\ \hat{x}_1^{\mathrm{in}} \\ \hat{p}_1^{\mathrm{in}} \\
    \vdots \\ \hat{x}_n^{\mathrm{in}} \\ \hat{p}_n^{\mathrm{in}}
  \end{bmatrix},
\end{equation}
where $\mat{A}$ is a $2(n+1)\times 2(n+1)$ matrix. To preserve
canonical commutator relations, $\mat{A}$ must be a symplectic matrix
\cite{Simon.PhysRevA.36.3868(1987)}, which we shall of course assume
to hold in the following. The first two rows of this set of equations
can be rewritten as:
\begin{equation}
\label{eq:Transform_Langevin}
  \begin{bmatrix}
    \Xout \\ \Pout
  \end{bmatrix}
  =
  \begin{bmatrix}
    A_{11} & A_{12} \\ A_{21} & A_{22}
  \end{bmatrix}
  \begin{bmatrix}
    \Xin \\ \Pin
  \end{bmatrix}
  +
  \begin{bmatrix}
    \hat{F}_x \\ \hat{F}_p
  \end{bmatrix},
\end{equation}
where, $\hat{\vec{F}} = [\hat{F}_x\:\hat{F}_p]^{\mathrm{T}}$, are
noise operators and represent the combined influence of the reservoir
modes.

Our second assumption about $E$ is that all the reservoir modes are
described by Gaussian states and are uncorrelated to the input state,
$\mean{\Xin\hat{F}_x} = \mean{\Xin}\mean{\hat{F}_x}$, etc. This means
in particular that the operators $\hat{\vec{F}}$ show Gaussian
fluctuations, and in order to preserve the commutation relation of the
output mode it is required that $[\hat{F}_x, \hat{F}_p] =
i(1-\mathrm{det}(\tilde{\mat{A}}))$, where $\tilde{\mat{A}}$ is the
upper $2\times 2$ block of $\mat{A}$ used in
Eq.~(\ref{eq:Transform_Langevin}). The second moments of the input
operators, the output operators, and the noise operators are all given
by covariance matrices. For instance, for the output mode the
covariance matrix reads $\CovMat_{\mathrm{out}} =
\mean{2\Re\{\delta\hat{\vec{y}}_{\mathrm{out}} \cdot
  \delta\hat{\vec{y}}_{\mathrm{out}}^{\mathrm{T}}\}}$, where the
vector $\hat{\vec{y}}_{\mathrm{out}} = [\Xout\:\Pout]^{\mathrm{T}}$
and where $\hat{\vec{y}}_{\mathrm{out}} =
\mean{\hat{\vec{y}}_{\mathrm{out}}} +
\delta\hat{\vec{y}}_{\mathrm{out}}$ defines the fluctuations of
operators around their mean values. The output mode covariance matrix
thus reads:
\begin{equation}
  \CovMat_{\mathrm{out}} = 2
  \begin{bmatrix}
    \Var(\Xout) & \Covar(\Xout,\Pout)\\
    \Covar(\Xout,\Pout) & \Var(\Pout)
  \end{bmatrix},
\end{equation}
where $\Covar(\Xout,\Pout) = \frac{1}{2}\mean{\delta\Xout\delta\Pout +
  \delta\Pout\delta\Xout}$. Similar covariance matrices
$\CovMat_{\mathrm{in}}$ and $\CovMat_F$ are defined for the input and
the noise parts, respectively. The second moments of the operators,
i.e., the covariance matrices, fulfill:
\begin{equation}
\label{eq:Transform_second_moments}
  \CovMat_{\mathrm{out}} = \tilde{\mat{A}}\CovMat_{\mathrm{in}}
     \tilde{\mat{A}}^{\mathrm{T}} + \CovMat_F.
\end{equation}
It was shown recently that coherent states suffice as input states to
fully characterize a process on harmonic oscillator modes
\cite{Lobino.Science.322.563(2008)}, and in the case of a Gaussian
process with the linear transformation
(\ref{eq:General_linear_transform}) of the canonical variables, a
small discrete set of coherent states is enough to yield the complete
information about the process
\cite{Wang.PhysRevA.88.022101(2013)}. Gaussian states are described
completely by their first and second moments, and the matrix
$\tilde{\mat{A}}$, the two mean values $\mean{\hat{F}_x}$ and
$\mean{\hat{F}_p}$, and the three real parameters of $\CovMat_{F}$ are
sufficient to describe the entire Gaussian process. We shall now show
how the process may equivalently be characterized by the quantities
indicated in Fig.~\ref{fig:BasicSetup}(a), which are experimentally
available when applying coherent input states to the process.

For the vacuum input state $\mean{\Xout}=\mean{\hat{F}_x}$ and
$\mean{\Pout}=\mean{\hat{F}_p}$ map out the mean values of the noise
operators of the environment, and then two other coherent input states
with non-zero mean values suffice to map out the entries of the matrix
$\tilde{\mat{A}}$, since $\mean{\vec{y}_{\mathrm{out}}} =
\tilde{\mat{A}}\mean{\vec{y}_{\mathrm{in}}} +
\mean{\hat{\vec{F}}}$. In turn, since $\CovMat_{\mathrm{in}}$ is the
identity matrix for any coherent state, the second moments of the
output mode operators establish the relations, $\Var(\hat{F}_x) =
\Var(\Xout) - \frac{1}{2}(A_{11}^2 + A_{12}^2)$, $\Var(\hat{F}_p) =
\Var(\Pout) - \frac{1}{2}(A_{21}^2 + A_{22}^2)$, and
$\Covar(\hat{F}_x,\hat{F}_p) = \Covar(\Xout,\Pout) -
\frac{1}{2}(A_{11}A_{21} + A_{12}A_{22})$. In the following, we assume
without loss of generality that $\mean{\hat{F}_x} = \mean{\hat{F}_p} =
0$, since any known non-zero mean value added to the output mode can
be readily identified by experiment and subtracted by a simple
displacement, which will not degrade our knowledge of the quantum
state. It is convenient to use the parametrization for the second
moments of the output mode shown in Fig.~\ref{fig:BasicSetup}(b),
i.e., the variances, $\sigma_1^2$ and $\sigma_2^2$, along the main
axes of the ``noise ellipse'' and the angle $\theta$ between the
$x$-axis and the major axis of the noise ellipse. These variables
relate to the parameters $\sigma_x^2 = \Var(\Xout)$, $\sigma_p^2 =
\Var(\Pout)$, and $C_{x,p} = \Covar(\Xout,\Pout)$ of
$\CovMat_{\mathrm{out}}$ by:
\begin{equation}
  \begin{split}
  \sigma_1^2 = \bar{\sigma}^2 + &\delta\sigma^2, \quad
  \sigma_2^2 = \bar{\sigma}^2 - \delta\sigma^2, \quad
  \tan(2\theta) = \frac{2C_{x,p}}{\sigma_x^2 - \sigma_p^2}, \\
  \text{with }\bar{\sigma}^2 &= \frac{\sigma_x^2 + \sigma_p^2}{2}, \quad
  \delta\sigma^2 = \sqrt{\frac{1}{4}(\sigma_x^2-\sigma_p^2)^2 + C_{x,p}^2}.
  \end{split}
\end{equation}
If $C_{x,p}$ is positive (negative), $0 < \theta <
\frac{\pi}{2}$ ($-\frac{\pi}{2} < \theta < 0$), while if $\sigma_x^2 =
\sigma_p^2$ and $C_{x,p} \ne 0$ we assume $\theta =
\frac{\pi}{4}\mathrm{sign}(C_{x,p}).$

For a coherent input state, $\densin = \ket{\alpha}\bra{\alpha}$, the
output state $\densout$ can always be described by a displaced,
squeezed, thermal state, offering enough variables to parametrize the
Gaussian state, illustrated in Fig.~\ref{fig:BasicSetup}(b). We shall
now provide a convenient expression of this output state as a function
of the coherent state amplitude, $\alpha$, and the parameters,
$A_{11}$, $A_{12}$, $A_{21}$, $A_{22}$, $\sigma_1^2$, $\sigma_2^2$,
and $\theta$, discussed above. To this end we define first the thermal
state:
\begin{equation}
\label{eq:dens_thermal_state}
  \dens_0 = \frac{1}{\pi\bar{n}_0}\int d^2\gamma
     e^{-|\gamma|^2/\bar{n}_0}\ket{\gamma}\bra{\gamma},
\end{equation}
where the integral is carried out over all coherent states
$\ket{\gamma}$.  Applying the squeezing operator $\hat{S}(r) =
e^{\frac{r}{2}(\a^2 - \a^{\dagger\,2})}$, where $r$ is a real
parameter, to the thermal state we obtain the squeezed thermal state:
$\densSTS = \hat{S}(r)\dens_0\hat{S}^{\dagger}(r)$ with well-known
properties \cite{Kim.PhysRevA.40.2494(1989)}. With the standard
definitions $\hat{X} = \frac{\a+\adag}{\sqrt{2}}$ and $\hat{P} =
\frac{-i(\a-\adag)}{\sqrt{2}}$ this state has $\Var(\hat{X}) =
(\bar{n}_0 + \frac{1}{2})e^{-2r}$ and $\Var(\hat{P}) = (\bar{n}_0 +
\frac{1}{2})e^{2r}$. By choosing appropriately the values of
$\bar{n}_0$ and $r$,
\begin{equation}
\label{eq:RelateVar_to_n_and_r}
  \sigma_1^2 = (\bar{n}_0 + \frac{1}{2})e^{-2r}, \quad
  \sigma_2^2 = (\bar{n}_0 + \frac{1}{2})e^{2r}.
\end{equation}
and applying, finally, the rotation operator
$\hat{R}(\theta) = e^{i\theta\adag\a}$ we obtain a rotated squeezed
thermal state,
\begin{equation}
\label{eq:rho-r}
   \dens_{\mathrm{r}} =
   \hat{R}(\theta)\densSTS\hat{R}^{\dagger}(\theta),
\end{equation}
with precisely the noise properties indicated by the output ellipse
shown in Fig.~\ref{fig:BasicSetup}(b).

The correct dependence of the output state mean values on the
amplitude of the input coherent state is reproduced by applying the
displacement operator $\hat{D}(\bar{\alpha}) =
e^{\bar{\alpha}\adag-\bar{\alpha}^*\a}$ to $\dens_{\mathrm{r}}$ such
that a coherent input state is mapped to the output state,
\begin{equation}
  E(|\alpha\rangle\langle\alpha|)= \dens_{\alpha},
\end{equation}
with
\begin{equation}
\label{eq:dens_alpha}
\begin{split}
  \dens_{\alpha} = &\hat{D}(\bar{\alpha})\hat{R}(\theta)\hat{S}(r)\dens_0
     \hat{S}^{\dagger}(r)\hat{R}^{\dagger}(\theta)\hat{D}^{\dagger}(\bar{\alpha}),\\
  &\begin{bmatrix}
    \bar{\alpha}_{\mathrm{R}} \\ \bar{\alpha}_{\mathrm{I}}
  \end{bmatrix}
  =
  \begin{bmatrix}
    A_{11} & A_{12} \\ A_{21} & A_{22}
  \end{bmatrix}
  \begin{bmatrix}
    \alpha_{\mathrm{R}} \\ \alpha_{\mathrm{I}}
  \end{bmatrix},
\end{split}
\end{equation}
where ``R'' and ``I'' refer to the real and imaginary parts,
respectively, of input mean amplitude $\alpha$ and output mean
amplitude $\bar{\alpha}$. It is convenient to introduce the equivalent
relations between $\alpha$ and $\bar{\alpha}$ in complex notation:
\begin{equation}
\label{eq:def_C_and_D}
  \begin{split}
    \bar{\alpha} &= C\alpha + D\alpha^*, \\
    C &= \frac{1}{2}(A_{11}-i A_{12} + iA_{21} + A_{22}), \\
    D &= \frac{1}{2}(A_{11}+i A_{12} + iA_{21} - A_{22}).
  \end{split}
\end{equation}
We note that rather than presenting a map on the input coherent state,
Eq.~(\ref{eq:dens_alpha}) formally provides the output state as an
$\alpha$-dependent transformation of a definite input state:
$|\alpha\rangle\langle\alpha| \rightarrow E(\ket{\alpha}\bra{\alpha})
\equiv
\hat{D}(\bar{\alpha})\dens_{r}\hat{D}^{\dagger}(\bar{\alpha})$. This
form is, however, perfectly useful to characterize the process and it
is a good starting point for our analysis of the qubit fidelity in the
next section.

\section{Derivation of the qubit fidelity formula}
\label{sec:QubitFidelity}

From the coherent-state expansion on the Fock-state basis,
\begin{equation}
  \ket{\alpha} = e^{-\frac{|\alpha|^2}{2}}\sum_{n=0}^{\infty}
     \frac{\alpha^n}{\sqrt{n!}}\ket{n},
\end{equation}
we see that the Fock basis states can be formally obtained from
expressions involving coherent states by $\ket{n} =
\frac{1}{\sqrt{n!}}  \frac{\partial^n}{\partial\alpha^n}
[e^{\frac{|\alpha|^2}{2}} \ket{\alpha}]|_{\alpha = 0}$. In turn, due
to the linearity of the map $E$, its action on a general Fock state
outer product can be retrieved as:
\begin{equation}
\label{eq:E(|n><m|)_formula}
   E(\ket{n}\bra{m}) = \frac{1}{\sqrt{n!m!}}
     \frac{\partial^n}{\partial\alpha^n}
     \frac{\partial^m}{\partial\alpha^{*m}}
     [e^{|\alpha|^2}E(\ket{\alpha}\bra{\alpha})]|_{\alpha=0}.
\end{equation}
Any qubit state expanded on the Fock states $|n=0\rangle$ and
$|n=1\rangle$ can thus be mapped if we know the quantities
$E(\ket{n=0}\bra{n=0})=E(\ket{\alpha=0}\bra{\alpha=0})$,
$E(\ket{1}\bra{0}) = \frac{\partial}{\partial\alpha}
E(\ket{\alpha}\bra{\alpha})|_{\alpha=0}$, $E(\ket{0}\bra{1}) =
\frac{\partial}{\partial\alpha^*}
E(\ket{\alpha}\bra{\alpha})|_{\alpha=0}$, and $E(\ket{1}\bra{1}) = (1
+ \frac{\partial^2}{\partial\alpha\partial\alpha^*})
E(\ket{\alpha}\bra{\alpha})|_{\alpha=0}$.

The derivatives can be expressed in
terms of $\bar{\alpha}$ using Eq.~(\ref{eq:def_C_and_D}):
\begin{align}
\notag
    \frac{\partial}{\partial\alpha} &= C\frac{\partial}{\partial\bar{\alpha}}
      + D^*\frac{\partial}{\partial\bar{\alpha}^*}, \\
    \frac{\partial}{\partial\alpha^*} &= D\frac{\partial}{\partial\bar{\alpha}}
      + C^*\frac{\partial}{\partial\bar{\alpha}^*}, \\
\notag
    \frac{\partial^2}{\partial\alpha\partial\alpha^*} &=
    CD\frac{\partial^2}{\partial\bar{\alpha}^2}
    +(|C|^2+|D|^2)\frac{\partial^2}{\partial\bar{\alpha}\partial\bar{\alpha}^*}
    + (CD)^*\frac{\partial^2}{\partial\bar{\alpha}^{*2}}.
\end{align}

Only the displacement operators in Eq.~(\ref{eq:dens_alpha}) depend on
the coherent state amplitudes, and their derivatives are given by
$\frac{\partial\hat{D}(\bar{\alpha})}{\partial\bar{\alpha}} =
\left(\adag - \frac{\bar{\alpha}^*}{2}\right)\hat{D}(\bar{\alpha})$,
$\frac{\partial\hat{D}(\bar{\alpha})}{\partial\bar{\alpha}^*} =
-\left(\a - \frac{\bar{\alpha}}{2}\right)\hat{D}(\bar{\alpha})$, and
their hermitian conjugates. The first and second derivatives of
$E(|\alpha\rangle\langle\alpha|)$ with respect to $\bar{\alpha}$ and
$\bar{\alpha}^*$ are thus given by
\begin{align}
\notag
  \frac{\partial E}{\partial\bar{\alpha}} &=
   \adag\dens_{\mathrm{r}} - \dens_{\mathrm{r}}\adag, \\
\notag
  \frac{\partial E}{\partial\bar{\alpha}^*} &=
   -\a\dens_{\mathrm{r}} + \dens_{\mathrm{r}}\a, \\
\label{eq:dE_dalpha_bar}
  \frac{\partial^2 E}{\partial\bar{\alpha}^2} &=
   \a^{\dagger\,2}\dens_{\mathrm{r}} - 2\adag\dens_{\mathrm{r}}\adag
   + \dens_{\mathrm{r}}\a^{\dagger\,2}, \\
\notag
  \frac{\partial^2E}{\partial\bar{\alpha}^{*2}} &=
   \a^2\dens_{\mathrm{r}} - 2\a\dens_{\mathrm{r}}\a + \dens_{\mathrm{r}}\a^2, \\
\notag
  \frac{\partial^2E}{\partial\bar{\alpha}\partial\bar{\alpha}^*}  &=
  -\adag\a\dens_{\mathrm{r}} + \adag\dens_{\mathrm{r}}\a
  + \a\dens_{\mathrm{r}}\adag - \dens_{\mathrm{r}}\a\adag,
\end{align}
where $\dens_{\mathrm{r}}$ is given in Eq.~(\ref{eq:rho-r}), and the
right hand sides are formally independent of $\alpha$ (the derivatives
are evaluated at $\alpha=0$).

The fidelity is defined as the overlap of the state subject to the
transformation $E$ with the original qubit state and thus requires
matrix elements of the left-hand side of
Eq.~(\ref{eq:E(|n><m|)_formula}) between the Fock states $\ket{0}$ and
$\ket{1}$. In turn, using
Eqs.~(\ref{eq:E(|n><m|)_formula})-(\ref{eq:dE_dalpha_bar}) this is
equivalent to calculating matrix element of the right-hand side of
Eq.~(\ref{eq:dE_dalpha_bar}) between the Fock states $\ket{0}$ and
$\ket{1}$. Now, due to the raising and lowering operators in this
equation (up to quadratic order) we end up with matrix elements on the
form $\bra{n'}\dens_{\mathrm{r}}\ket{m'} =
e^{i\theta(n'-m')}\bra{n'}\densSTS\ket{m'}$, where the integers $n'$
and $m'$ may take values from 0 to 3. For instance, we have
$\bra{1}E(\ket{1}\bra{0})\ket{0} = C\bra{1}\adag\dens_{\mathrm{r}} -
\dens_{\mathrm{r}}\adag\ket{0} + D^*\bra{1}-\a\dens_{\mathrm{r}} +
\dens_{\mathrm{r}}\a\ket{0} = C[\bra{0}\densSTS\ket{0}
-\bra{1}\densSTS\ket{1}] -
\sqrt{2}D^*e^{2i\theta}\bra{2}\densSTS\ket{0}$, and the first term in
this expression can be calculated directly as
\begin{align}
\notag
  \bra{0}\densSTS\ket{0}
   &= \frac{1}{\pi\bar{n}_0}\int d^2\gamma e^{-|\gamma|^2/\bar{n}_0}
     |\bra{0}\hat{S}(r)\ket{\gamma}|^2 \\
\notag
   &= \frac{1}{\pi\bar{n}_0\cosh(r)}\int d^2\gamma
    e^{-\frac{1+\bar{n}_0}{\bar{n}_0}|\gamma|^2 + \frac{\gamma^2 + \gamma^{*2}}{2}\tanh(r)} \\
\notag
   &= \frac{1}{\sqrt{\left[(\frac{1}{2}+\bar{n}_0)e^{-2r}+\frac{1}{2}\right]
      \left[(\frac{1}{2}+\bar{n}_0)e^{2r}+\frac{1}{2}\right]}} \\
   &= \frac{1}{[(\sigma_1^2+\frac{1}{2})(\sigma_2^2+\frac{1}{2})]^{1/2}}.
\label{eq:<0|densSTS|0>_example}
\end{align}
The first equality, in which $\bra{n=0}$ refers to the Fock basis and
$\ket{\gamma}$ to the coherent-state basis, follows from the expansion
~(\ref{eq:dens_thermal_state}) of the thermal state on coherent
states, the second line exploits the Fock-state expansion of squeezed
coherent states \cite{Gong.AmJPhys.58.1003(1990)}:
\begin{equation}
\label{eq:SqueezedCohState_FockBasis}
  \begin{split}
  \bra{n}\hat{S}(r)\ket{\gamma} =
   &\frac{e^{-\frac{|\gamma|^2}{2} + \frac{\gamma^2}{2}\tanh(r)}}{\sqrt{n!\cosh(r)}}
    \left(\frac{1}{2}\tanh(r)\right)^{\frac{n}{2}} \\
   &\times H_n\left(\gamma/\sqrt{\sinh(2r)}\right),
  \end{split}
\end{equation}
where $H_n$ is a Hermite polynomial, the third line carries out the
$\gamma$-integration, and the last step applies the relations in
Eq.~(\ref{eq:RelateVar_to_n_and_r}). Similar calculations are readily
performed for the remaining relevant matrix elements and yield:
\begin{equation}
\label{eq:<a|densSTS|b>_remaining}
  \begin{split}
  \bra{1}\densSTS\ket{1} &= \frac{\sigma_1^2\sigma_2^2-\frac{1}{4}}
    {[(\sigma_1^2+\frac{1}{2})(\sigma_2^2+\frac{1}{2})]^{3/2}}, \\
  \bra{0}\densSTS\ket{2} &= \frac{\sigma_1^2-\sigma_2^2}
    {2\sqrt{2}[(\sigma_1^2+\frac{1}{2})(\sigma_2^2+\frac{1}{2})]^{3/2}}, \\
  \bra{2}\densSTS\ket{2} &= \frac{\left(\sigma_1^2\sigma_2^2
     -\frac{1}{4}\right)^2 + \frac{1}{8}(\sigma_1^2-\sigma_2^2)^2}
  {[(\sigma_1^2+\frac{1}{2})(\sigma_2^2+\frac{1}{2})]^{5/2}}\\
  \bra{1}\densSTS\ket{3} &= \frac{\sqrt{6}(\sigma_1^2\sigma_2^2 - \frac{1}{4})
    (\sigma_1^2 - \sigma_2^2)}
   {4[(\sigma_1^2+\frac{1}{2})(\sigma_2^2+\frac{1}{2})]^{5/2}}.
  \end{split}
\end{equation}
By integrating the fidelity for any input qubit state,
$\ket{\psi(\Omega)} = \cos\frac{\theta}{2}\ket{0} +
e^{i\phi}\sin\frac{\theta}{2}\ket{1}$ with $0 \le \theta \le \pi$ and
$0 \le \phi \le 2\pi$, we determine the average qubit fidelity $\Fq$ :
\begin{equation}
\label{eq:Fq_first_principles}
  \begin{split}
  \Fq &= \frac{1}{4\pi}\int d\Omega \bra{\psi(\Omega)}
    E(\ket{\psi(\Omega)}\bra{\psi(\Omega)})\ket{\psi(\Omega)} \\
   &= \frac{1}{3}[\bra{0}E(\ket{0}\bra{0})\ket{0}
                   +\bra{1}E(\ket{1}\bra{1})\ket{1}] \\
   &\quad +\frac{1}{6}[\bra{0}E(\ket{0}\bra{1})\ket{1} +
                 \bra{1}E(\ket{1}\bra{0})\ket{0}] \\
   &\quad +\frac{1}{6}[\bra{1}E(\ket{0}\bra{0})\ket{1}
              +\bra{0}E(\ket{1}\bra{1})\ket{0}],
  \end{split}
\end{equation}
where $\ket{0}$ and $\ket{1}$ refer to Fock states.
With the expression derived above, we thus reach the final, explicit
expression for the average qubit fidelity in terms of the mapping
parameters of the Gaussian process:
\begin{widetext}
\begin{equation}
\label{eq:Fq_arbitrary_A_no_rot}
  \begin{split}
  \Fq &=  \frac{1}{6\sqrt{(\sigma_1^2+\frac{1}{2})
    (\sigma_2^2+\frac{1}{2})}}\left\{3
   + \frac{3(\sigma_1^2\sigma_2^2-\frac{1}{4})}{(\sigma_1^2+\frac{1}{2})
    (\sigma_2^2+\frac{1}{2})} +
    \frac{\Re\{C + \tilde{D}^*\}}{\sigma_1^2+\frac{1}{2}}
   +\frac{\Re\{C-\tilde{D}^*\}}{\sigma_2^2+\frac{1}{2}}
   \right. \\ &\left.\quad
  - \frac{|C+\tilde{D}^*|^2(\sigma_1^2-1)}{(\sigma_1^2+\frac{1}{2})^2}
  - \frac{|C-\tilde{D}^*|^2(\sigma_2^2-1)}{(\sigma_2^2+\frac{1}{2})^2}
  -\frac{|C+\tilde{D}^*|^2(\sigma_2^2-\frac{1}{2}) +
   |C-\tilde{D}^*|^2(\sigma_1^2-\frac{1}{2})}
  {2 (\sigma_1^2+\frac{1}{2})(\sigma_2^2+\frac{1}{2})} \right\},
  \end{split}
\end{equation}
\end{widetext}
where $\tilde{D} = De^{-2i\theta}$. This is the main results of the article, and in the next section we
shall consider the fidelity formula in various specific cases,
corresponding to the experimental storage and transfer schemes
mentioned in the Introduction. 

Let us briefly discuss the different
effects contributing to a reduction of the fidelity.  First, we
observe that Eq.~(\ref{eq:Fq_arbitrary_A_no_rot}) decreases when
$\sigma_{1,2}$ become large. This is natural, as the qubit occupies
only the lowest two Fock states, while the output state is distributed
toward higher number states $n\propto \sigma_1^2, \sigma_2^2$, and
hence a corresponding smaller fraction of the population remains in
the qubit space. Even with $\sigma_{1,2}$ close to the minimum allowed
by the Heisenberg uncertainty relation, the values of $C, D$ and
$\theta$ can lead to large variations in the qubit fidelity. This is
associated with the possibility for the map to yield an (undesired)
unitary operation on the qubit, e.g., in the form of a rotation of the
Bloch vector around the $z$-axis, caused by a rotation of the
continuous quadrature variables in the $(\hat{X},\hat{P})$ phase
space. Thus, the unitary mapping $\hat{X}_1 \rightarrow -\hat{X}_1$
and $\hat{P}_1 \rightarrow -\hat{P}_1$, represented by $A_{11} =
A_{22} = -1$, $A_{12} = A_{21} = 0$, and $\sigma_1^2 = \sigma_2^2 =
\frac{1}{2}$, yields, according to
Eq.~(\ref{eq:Fq_arbitrary_A_no_rot}), an average qubit fidelity $\Fq
= \frac{1}{3}$. The mapping, however, is perfect, if we only redefine the basis states by a simple phase change of
$-\pi$ after the process, and it makes sense to allow incorporation of
such a trivial transformation in the definition of the average qubit
fidelity. The effect on Eq.~(\ref{eq:Fq_arbitrary_A_no_rot}) of a phase
rotation by $\theta'$ corresponds to setting $C \rightarrow
Ce^{i\theta'}$ and $\tilde{D}^* \rightarrow \tilde{D}^* e^{i\theta'}$,
which affects only the two terms linear in $C$ and $\tilde{D}^*$ in
Eq.~(\ref{eq:Fq_arbitrary_A_no_rot}). The angles $\theta'$ yielding
the extremal values of $\Fq$ are thus given by:
\begin{equation}
  e^{2i\theta'} = \frac{(C^*+\tilde{D})(\sigma_2^2+\frac{1}{2}) +
     (C^*-\tilde{D})(\sigma_1^2+\frac{1}{2})}
   {(C+\tilde{D}^*)(\sigma_2^2+\frac{1}{2}) +
    (C-\tilde{D}^*)(\sigma_1^2+\frac{1}{2})}.
\end{equation}
For the simple map with the $\pi$ rotation, qubit fidelity extrema are
found at $\theta' = n\pi$, where $n$ is an integer, and for odd $n$
the rotation is counter-acted and a unit fidelity is recovered.

\section{Examples}
\label{sec:Specific_examples}

\subsection{Symmetric gain and variance}
\label{sec:Symmetric_case}
\begin{figure}[t]
  \centering
  \includegraphics[width=\linewidth]{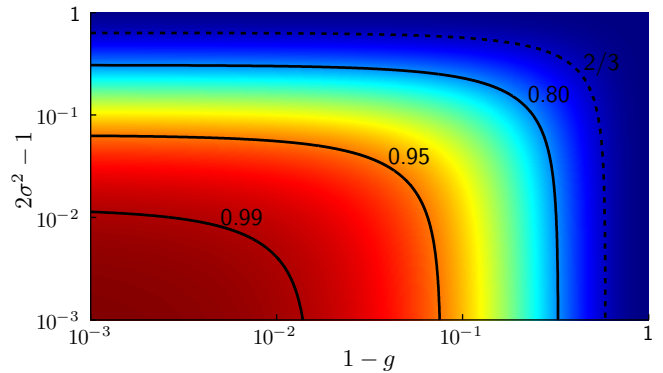}
  \caption{(Color online) A contour plot of the qubit fidelity for
    symmetric gain and variances according to
    Eq.~(\ref{eq:Fq_symmetric_gain_var}) as a function of the gain
    imperfection, $1-g$, and the excess variance relative to the
    vacuum noise limit, $2\sigma^2-1$. A few values of
    $\Fq$ are marked on the graph with the dashed curve
    enclosing the non-classical limit $\Fq > \frac{2}{3}$.}
  \label{fig:Fidelity_sym_gain_and_var}
\end{figure}
Consider the specific case where both $\hat{X}$ and $\hat{P}$ are
multiplied by the same gain coefficient $g$ in the transformation
process such that $A_{11} = A_{22} = g$ and $A_{12} = A_{21} =
0$. Assume also the added noise to be symmetric, $\sigma_1^2 =
\sigma_2^2 = \sigma^2$. Then the fidelity becomes
\begin{equation}
\label{eq:Fq_symmetric_gain_var}
  \Fq = \frac{6\sigma^4 + 3\sigma^2 + g(2\sigma^2+1)
    - g^2(3\sigma^2-\frac{5}{2})}{6(\sigma^2+\frac{1}{2})^3},
\end{equation}
which is identical to the result found in
\cite{Sherson.Nature.443.557(2006)}. The value of $\Fq$ as a function
of $g$ and $\sigma^2$ is shown in
Fig.~\ref{fig:Fidelity_sym_gain_and_var}.

By a projective qubit measurement, one obtains an outcome that may be
stored by classical means, and the corresponding eigenstate may be
reinstalled in the physical output system at any later time. This
classical procedure provides a qubit state with an average overlap
with the unknown initial state of $2/3$
\cite{Bouwmeester.JModOpt.47.279(2000)}. The dashed curve with
$F_q=2/3$ in Fig.~\ref{fig:Fidelity_sym_gain_and_var} represents the benchmark value where a quantum storage
or transfer operation outperforms the much simpler classical strategy.

\subsection{An oscillator coupled to a heat bath}
\label{sec:Osc_to_heat_bath}
\begin{figure}[t]
  \centering
  \includegraphics[width=\linewidth]{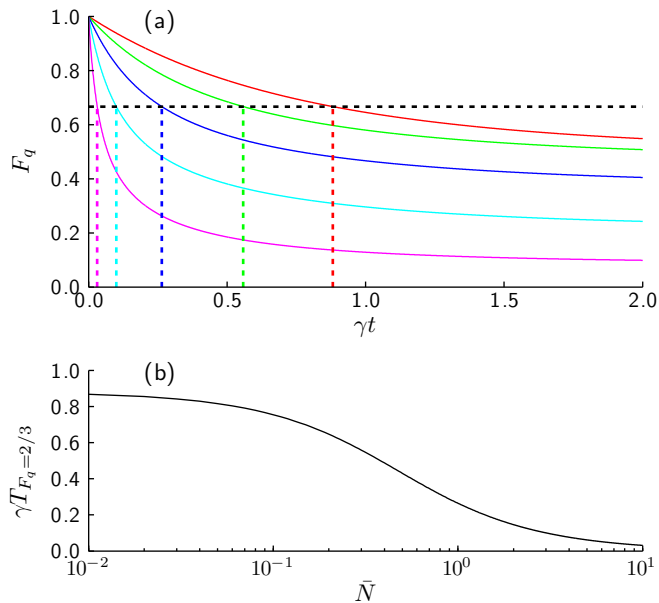}
  \caption{(Color online) (a) The decay of qubit fidelity when the
    harmonic oscillator hosting the qubit is coupled to a heat bath by
    a decay rate $\gamma$. Each curve corresponds to a specific bath
    temperature with $\bar{N}$ denoting the mean excitation level of
    the oscillator in equilibrium. From above: $\bar{N} = 0$ (red),
    $\bar{N} = 0.3$ (green), $\bar{N} = 1$ (blue), $\bar{N} = 3$
    (cyan), and $\bar{N} = 10$ (magenta). The horizontal dashed line
    denotes the non-classical limit $\Fq > \frac{2}{3}$ and the
    vertical dashed lines mark the time $T_{\Fq = 2/3}$ at which this
    limit is reached. This characteristic time is also shown in (b) on
    the vertical axis (in units of $\gamma^{-1}$) as a function of the
    equilibrium excitation level $\bar{N}$.}
  \label{fig:Fidelity_heat_bath}
\end{figure}
Consider a harmonic oscillator, e.g.~a cavity field with resonance
frequency $\omega_0$, coupled by an energy-decay rate $\gamma$ to an
external heat bath at temperature $T$. The characteristic number of
excitations in the heat bath is $\bar{N} =
[\exp(\frac{\hbar\omega_0}{k_{\mathrm{B}}T})-1]^{-1}$, and the quantum
Langevin equations for the oscillator mode $\a$ can be written
\cite{Gardiner.PhysRevA.31.3761(1985)}:
\begin{equation}
\label{eq:Fq_symmetric}
  \frac{\partial\a}{\partial t} = -i\omega_0\a - \frac{\gamma}{2}\a
   - \sqrt{\gamma}\bin,
\end{equation}
where $\bin$ is the input thermal field, which in the broad-band
approximation satisfies $\mean{\bin(t)\bin^{\dagger}(t')} =
(\bar{N}+1)\delta(t-t')$ and $\mean{\bin^{\dagger}(t)\bin(t')} =
\bar{N}\delta(t-t')$.  Eq.~(\ref{eq:Transform_Langevin}) yields the
solution of Eq.~(\ref{eq:Fq_symmetric}) with $g \equiv A_{11} = A_{22}
= e^{-\frac{\gamma t}{2}}$ and $A_{12} = A_{21} = 0$. From the
properties of $\bin$ we deduce that $\Var(\hat{F}_x) = \Var(\hat{F}_p)
= (\bar{N}+\frac{1}{2})(1-e^{-\gamma t})$ and
$\Covar(\hat{F}_x,\hat{F}_p) = 0$, and hence for a coherent-state
input the variances of the output state is $\sigma^2 \equiv \sigma_1^2
= \sigma_2^2 = \frac{1}{2} + \bar{N}(1-e^{-\gamma t})$. The qubit
fidelity now follows from inserting the parameters $g$ and $\sigma^2$
into Eq.~(\ref{eq:Fq_symmetric_gain_var}), and the resulting fidelity
is shown in Fig.~\ref{fig:Fidelity_heat_bath}.

We observe that the decay of fidelity occurs faster when the heat bath
temperature is increased. In Fig.~\ref{fig:Fidelity_heat_bath}(a) the
initial linear decrease in $\Fq$ follows the approximate formula: $\Fq
\approx 1 - \frac{(2+5\bar{N})\gamma t}{3}$. In the asymptotic limit
$t \rightarrow \infty$ the fidelity converges, $\Fq \rightarrow
\frac{\bar{N}+\frac{1}{2}}{(\bar{N}+1)^2}$, i.e.~for $\bar{N} = 0$ the
qubit decays to the ground state $\ket{0}$ which has a 50\:\% chance
of reproducing the random input qubit, and for large $\bar{N}$ the
oscillator is most likely excited away from the qubit space spanned by
$\ket{0}$ and $\ket{1}$ leading to a vanishing fidelity.

The horizontal dashed line with $F_q=2/3$ in
Fig.~\ref{fig:Fidelity_heat_bath}(a) represents the benchmark value of
quantum storage, which occurs at the $\bar{N}$-dependent times marked
by the vertical dashed lines. In Fig.~\ref{fig:Fidelity_heat_bath}(b)
the these times are shown more generally as a function of
$\bar{N}$. The $\bar{N} \rightarrow 0$ limit yields $\gamma
T_{\Fq=2/3} \rightarrow -\ln(\sqrt{2}-1) \approx 0.88$, i.e.~for an
exponentially decreasing coherence, the process supersedes the
classical benchmark for times less than 88\:\% of the coherence time.

\subsection{Asymmetric gain and variance along the same major axes}
\label{sec:Asym_same_major_axes}
\begin{figure}[t]
  \centering
  \includegraphics[width=\linewidth]{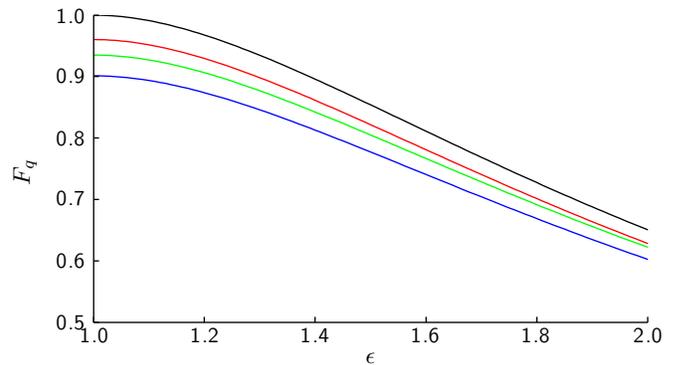}
  \caption{(Color online) The qubit fidelity from
    Eq.~(\ref{eq:Fq_asym}) as a function of $\epsilon$, which is
    conveniently used to parametrize the asymmetry in gain and
    variance by $g_x = g_0\epsilon$, $g_p = g_0/\epsilon$, $\sigma_x^2
    = \sigma_0^2\epsilon^2$, and $\sigma_p^2 =
    \sigma_0^2/\epsilon^2$. From the top, $g_0 = 1$ and $\sigma_0^2 =
    \frac{1}{2}$ (black), $g_0 = 1$ and $\sigma_0^2 = \frac{1.05}{2}$
    (red), $g_0 = 0.9$ and $\sigma_0^2 = \frac{1}{2}$ (green), and
    $g_0 = 0.9$ and $\sigma_0^2 = \frac{1.05}{2}$ (blue).}
  \label{fig:Fidelity_asym_gain_and_var}
\end{figure}
In most practical cases with asymmetric gain and variance, the
asymmetries materialize along the same axes in
$(\hat{X},\hat{P})$-space. One example is the degenerate parametric
amplifier \cite{Gardiner.QuantumNoise}, for which the transformations
are $\mean{\Xout} = G\mean{\Xin}$, $\mean{\Pout} = G^{-1}\mean{\Pin}$,
$\sigma_x^2 = \frac{G^2}{2}$, $\sigma_p^2 = \frac{1}{2G^2}$, and
$C_{x,p} = 0$, i.e.~the coordinate system is chosen, without loss of
generality, such that the mean value transformation $\tilde{\mat{A}}$
is diagonal and at the same time it turns out that $\theta = 0$,
i.e.~the $(\hat{X},\hat{P})$-axes form also the major axes for the
covariance matrix $\CovMat_{\mathrm{out}}$. Another example can be
found in spin-ensemble based quantum memories, which encode quantum
information into the transverse components $\hat{X} \equiv
\hat{S}_x/\sqrt{|S_z|}$ and $\hat{P} \equiv -\hat{S}_y/\sqrt{|S_z|}$
of a macroscopic spin polarized along the negative $z$-direction. For
an inhomogeneous distribution of spin frequencies the stored
information is ``diffused'' into the spin ensemble and recalled as a
spin echo using a set of $\pi$ pulses for inverting the ensemble
population. These $\pi$ pulses employ a spin rotation around a certain
axis and thereby break the symmetry of the $(\hat{X},\hat{P})$-space,
and especially for non-ideal $\pi$ pulses the
transformations~(\ref{eq:Transform_Langevin})
and~(\ref{eq:Transform_second_moments}) become asymmetric (in some
cases even squeezed) \cite{Julsgaard.arXiv.1309.5517(2013)}. In this
case also, $\tilde{\mat{A}}$ and $\CovMat$ turn out diagonal in a
common coordinate system, and the two above examples thus motivate a
closer look on the particular transformation:
\begin{equation}
\label{eq:Transform_asym_same_axes}
  \tilde{\mat{A}} =
  \begin{bmatrix}
    g_x & 0 \\ 0 & g_p
  \end{bmatrix},
  \qquad
  \CovMat_{\mathrm{out}} =
  \begin{bmatrix}
    2\sigma_x^2 & 0 \\ 0 & 2\sigma_p^2
  \end{bmatrix}.
\end{equation}
As long as the Heisenberg uncertainty relation, $\sigma_x^2\sigma_p^2
\ge \frac{1}{4}$, is satisfied we allow $\sigma_x^2$ and $\sigma_p^2$
to take any value meeting the constraints $2\sigma_1^2 \ge g_x^2$ and
$2\sigma_2^2 \ge g_p^2$ imposed by
Eq.~(\ref{eq:Transform_second_moments}) and the positivity of
$(\gamma_F)_{11}$ and $(\gamma_F)_{22}$. We may think of this
transformations as a noisy parametric amplifier (the version discussed
above is a minimum-uncertainty case). When the properties
of~(\ref{eq:Transform_asym_same_axes}) are inserted into the general
formula~(\ref{eq:Fq_arbitrary_A_no_rot}) we find:
\begin{equation}
  \label{eq:Fq_asym}
  \begin{split}
    \Fq &= \frac{1}{6\sqrt{(\sigma_x^2+\frac{1}{2})
     (\sigma_p^2+\frac{1}{2})}}\left\{3 +
     \frac{3(\sigma_x^2\sigma_p^2-\frac{1}{4})}{(\sigma_x^2+\frac{1}{2})
     (\sigma_p^2+\frac{1}{2})} \right. \\
  &\quad +  \frac{g_x}{\sigma_x^2+\frac{1}{2}}
  +  \frac{g_p}{\sigma_p^2+\frac{1}{2}}
   -\frac{g_x^2(\sigma_x^2-1)}{(\sigma_x^2+\frac{1}{2})^2}
   - \frac{g_p^2(\sigma_p^2-1)}{(\sigma_p^2+\frac{1}{2})^2} \\
   &\quad  \left. - \frac{g_x^2(\sigma_p^2-\frac{1}{2})
     + g_p^2(\sigma_x^2-\frac{1}{2})}
    {2(\sigma_x^2+\frac{1}{2})(\sigma_p^2+\frac{1}{2})}\right\}.
  \end{split}
\end{equation}
In order to illustrate how the asymmetry affects the qubit fidelity,
we show in Fig.~\ref{fig:Fidelity_asym_gain_and_var} a number of
curves, where for each curve the products $g_xg_p \equiv g_0^2$ and
$\sigma_x^2\sigma_p^2 \equiv \sigma_0^4$ remain constant but the
degree of asymmetry is changed along the horizontal axis, see the
figure caption for explanation. We note that the upper curve
corresponds to the special case of a noiseless parametric amplifier
for which $\Fq =
\frac{\sqrt{2}}{3}\frac{\cosh(2r)+2\cosh(r)+3}{[1+\cosh(2r)]^{3/2}}$,
where we parametrized the gain as $G = \epsilon = e^r$. This
expression for $F_q$ stays above the classical benchmark $\frac{2}{3}$
for $\epsilon \lesssim 1.96$.

\section{Summary}
\label{sec:summary}

In this paper we have presented calculations yielding the average
fidelity for storage and transfer of qubit states which are encoded in
the $\ket{0}$ and $\ket{1}$ Fock states of a harmonic oscillator,
subjected to a Gaussian process. Since coherent states form a complete
basis for the harmonic oscillator, the parameters characterizing a
Gaussian process can be determined by its action on coherent states,
and subsequently the action of the process on any class of quantum
states can be obtained. The main result of our calculation is the
explicit expression, Eq.~(\ref{eq:Fq_arbitrary_A_no_rot}), for the
average qubit fidelity for a general Gaussian process. This expression
shows how imperfect gain and added noise both contribute to the
infidelity of protocols handling qubits in oscillator degrees of
freedom. It also shows, however, that part of the infidelity may be
recovered by merely redefining the phases of the qubit basis states.

There has already been considerable efforts to determine the fidelity
of Gaussian operations acting on oscillators prepared in coherent
states, squeezed states and qubit states, and in connection with the
beam-splitter like coupling of light modes and atomic ensembles, the
average qubit fidelity has been calculated in
Ref.~\cite{Sherson.Nature.443.557(2006)}. Our theory, indeed,
reproduces that result when we restrict to symmetric gain and
noise. Currently, however, there is a growing experimental interest in hybrid
quantum system architectures, where, e.g., effective two-level systems
are used for preparation and processing of qubit states, while
oscillator systems are used for storage and transport. These systems
apply different coupling schemes and frequently the couplings to the
quadratures of the electromagnetic, mechanical or collective
spin oscillators differ, leading to asymmetries in the
$(\hat{X},\hat{P})$ phase-space. Another source of asymmetry may occur
during processing of the individual oscillator modes, as exemplified
by $\pi$ pulses applied to spin ensembles in
Ref.~\cite{Julsgaard.arXiv.1309.5517(2013)}. The general expression
Eq.~(\ref{eq:Fq_arbitrary_A_no_rot}) and the examples given in
Sec.~\ref{sec:Specific_examples} properly
describe the fidelity of qubit manipulation in such hybrid systems.

\begin{acknowledgments}
  The authors acknowledge support from the EU Seventh Framework
  Programme collaborative project iQIT and the Villum
  Foundation.
\end{acknowledgments}

%\bibliography{bibfile}

\begin{thebibliography}{45}%
\makeatletter
\providecommand \@ifxundefined [1]{%
 \@ifx{#1\undefined}
}%
\providecommand \@ifnum [1]{%
 \ifnum #1\expandafter \@firstoftwo
 \else \expandafter \@secondoftwo
 \fi
}%
\providecommand \@ifx [1]{%
 \ifx #1\expandafter \@firstoftwo
 \else \expandafter \@secondoftwo
 \fi
}%
\providecommand \natexlab [1]{#1}%
\providecommand \enquote  [1]{``#1''}%
\providecommand \bibnamefont  [1]{#1}%
\providecommand \bibfnamefont [1]{#1}%
\providecommand \citenamefont [1]{#1}%
\providecommand \href@noop [0]{\@secondoftwo}%
\providecommand \href [0]{\begingroup \@sanitize@url \@href}%
\providecommand \@href[1]{\@@startlink{#1}\@@href}%
\providecommand \@@href[1]{\endgroup#1\@@endlink}%
\providecommand \@sanitize@url [0]{\catcode `\\12\catcode `\$12\catcode
  `\&12\catcode `\#12\catcode `\^12\catcode `\_12\catcode `\%12\relax}%
\providecommand \@@startlink[1]{}%
\providecommand \@@endlink[0]{}%
\providecommand \url  [0]{\begingroup\@sanitize@url \@url }%
\providecommand \@url [1]{\endgroup\@href {#1}{\urlprefix }}%
\providecommand \urlprefix  [0]{URL }%
\providecommand \Eprint [0]{\href }%
\providecommand \doibase [0]{http://dx.doi.org/}%
\providecommand \selectlanguage [0]{\@gobble}%
\providecommand \bibinfo  [0]{\@secondoftwo}%
\providecommand \bibfield  [0]{\@secondoftwo}%
\providecommand \translation [1]{[#1]}%
\providecommand \BibitemOpen [0]{}%
\providecommand \bibitemStop [0]{}%
\providecommand \bibitemNoStop [0]{.\EOS\space}%
\providecommand \EOS [0]{\spacefactor3000\relax}%
\providecommand \BibitemShut  [1]{\csname bibitem#1\endcsname}%
\let\auto@bib@innerbib\@empty
%</preamble>
\bibitem [{\citenamefont
  {DiVincenzo}(1995)}]{DiVincenzo.Science.270.255(1995)}%
  \BibitemOpen
  \bibfield  {author} {\bibinfo {author} {\bibfnamefont {D.~P.}\ \bibnamefont
  {DiVincenzo}},\ }\href {\doibase 10.1126/science.270.5234.255} {\bibfield
  {journal} {\bibinfo  {journal} {Science}\ }\textbf {\bibinfo {volume}
  {270}},\ \bibinfo {pages} {255} (\bibinfo {year} {1995})}\BibitemShut
  {NoStop}%
\bibitem [{\citenamefont {Boozer}\ \emph {et~al.}(2007)\citenamefont {Boozer},
  \citenamefont {Boca}, \citenamefont {Miller}, \citenamefont {Northup},\ and\
  \citenamefont {Kimble}}]{Boozer.PhysRevLett.98.193601(2007)}%
  \BibitemOpen
  \bibfield  {author} {\bibinfo {author} {\bibfnamefont {A.~D.}\ \bibnamefont
  {Boozer}}, \bibinfo {author} {\bibfnamefont {A.}~\bibnamefont {Boca}},
  \bibinfo {author} {\bibfnamefont {R.}~\bibnamefont {Miller}}, \bibinfo
  {author} {\bibfnamefont {T.~E.}\ \bibnamefont {Northup}}, \ and\ \bibinfo
  {author} {\bibfnamefont {H.~J.}\ \bibnamefont {Kimble}},\ }\href {\doibase
  10.1103/PhysRevLett.98.193601} {\bibfield  {journal} {\bibinfo  {journal}
  {Phys. Rev. Lett.}\ }\textbf {\bibinfo {volume} {98}},\ \bibinfo {pages}
  {193601} (\bibinfo {year} {2007})}\BibitemShut {NoStop}%
\bibitem [{\citenamefont {Specht}\ \emph {et~al.}(2011)\citenamefont {Specht},
  \citenamefont {N\"olleke}, \citenamefont {Reiserer}, \citenamefont {Uphoff},
  \citenamefont {Figueroa}, \citenamefont {Ritter},\ and\ \citenamefont
  {Rempe}}]{Specht.Nature.473.190(2011)}%
  \BibitemOpen
  \bibfield  {author} {\bibinfo {author} {\bibfnamefont {H.~P.}\ \bibnamefont
  {Specht}}, \bibinfo {author} {\bibfnamefont {C.}~\bibnamefont {N\"olleke}},
  \bibinfo {author} {\bibfnamefont {A.}~\bibnamefont {Reiserer}}, \bibinfo
  {author} {\bibfnamefont {M.}~\bibnamefont {Uphoff}}, \bibinfo {author}
  {\bibfnamefont {E.}~\bibnamefont {Figueroa}}, \bibinfo {author}
  {\bibfnamefont {S.}~\bibnamefont {Ritter}}, \ and\ \bibinfo {author}
  {\bibfnamefont {G.}~\bibnamefont {Rempe}},\ }\href {\doibase
  10.1038/nature09997} {\bibfield  {journal} {\bibinfo  {journal} {Nature}\
  }\textbf {\bibinfo {volume} {473}},\ \bibinfo {pages} {190} (\bibinfo {year}
  {2011})}\BibitemShut {NoStop}%
\bibitem [{\citenamefont {Riebe}\ \emph {et~al.}(2004)\citenamefont {Riebe},
  \citenamefont {H\"affner}, \citenamefont {Roos}, \citenamefont {H\"ansel},
  \citenamefont {Benhelm}, \citenamefont {Lancaster}, \citenamefont {K\"orber},
  \citenamefont {Becher}, \citenamefont {Schmidt-Kaler}, \citenamefont
  {James},\ and\ \citenamefont {Blatt}}]{Riebe.Nature.429.734(2004)}%
  \BibitemOpen
  \bibfield  {author} {\bibinfo {author} {\bibfnamefont {M.}~\bibnamefont
  {Riebe}}, \bibinfo {author} {\bibfnamefont {H.}~\bibnamefont {H\"affner}},
  \bibinfo {author} {\bibfnamefont {C.~F.}\ \bibnamefont {Roos}}, \bibinfo
  {author} {\bibfnamefont {W.}~\bibnamefont {H\"ansel}}, \bibinfo {author}
  {\bibfnamefont {J.}~\bibnamefont {Benhelm}}, \bibinfo {author} {\bibfnamefont
  {G.~P.~T.}\ \bibnamefont {Lancaster}}, \bibinfo {author} {\bibfnamefont
  {T.~W.}\ \bibnamefont {K\"orber}}, \bibinfo {author} {\bibfnamefont
  {C.}~\bibnamefont {Becher}}, \bibinfo {author} {\bibfnamefont
  {F.}~\bibnamefont {Schmidt-Kaler}}, \bibinfo {author} {\bibfnamefont
  {D.~F.~V.}\ \bibnamefont {James}}, \ and\ \bibinfo {author} {\bibfnamefont
  {R.}~\bibnamefont {Blatt}},\ }\href {\doibase 10.1038/nature02570} {\bibfield
   {journal} {\bibinfo  {journal} {Nature}\ }\textbf {\bibinfo {volume}
  {429}},\ \bibinfo {pages} {734} (\bibinfo {year} {2004})}\BibitemShut
  {NoStop}%
\bibitem [{\citenamefont {Barrett}\ \emph {et~al.}(2004)\citenamefont
  {Barrett}, \citenamefont {Chiaverini}, \citenamefont {Schaetz}, \citenamefont
  {Britton}, \citenamefont {Itano}, \citenamefont {Jost}, \citenamefont
  {Knill}, \citenamefont {Langer}, \citenamefont {Leibfried}, \citenamefont
  {Ozeri},\ and\ \citenamefont {Wineland}}]{Barrett.Nature.429.737(2004)}%
  \BibitemOpen
  \bibfield  {author} {\bibinfo {author} {\bibfnamefont {M.~D.}\ \bibnamefont
  {Barrett}}, \bibinfo {author} {\bibfnamefont {J.}~\bibnamefont {Chiaverini}},
  \bibinfo {author} {\bibfnamefont {T.}~\bibnamefont {Schaetz}}, \bibinfo
  {author} {\bibfnamefont {J.}~\bibnamefont {Britton}}, \bibinfo {author}
  {\bibfnamefont {W.~M.}\ \bibnamefont {Itano}}, \bibinfo {author}
  {\bibfnamefont {J.~D.}\ \bibnamefont {Jost}}, \bibinfo {author}
  {\bibfnamefont {E.}~\bibnamefont {Knill}}, \bibinfo {author} {\bibfnamefont
  {C.}~\bibnamefont {Langer}}, \bibinfo {author} {\bibfnamefont
  {D.}~\bibnamefont {Leibfried}}, \bibinfo {author} {\bibfnamefont
  {R.}~\bibnamefont {Ozeri}}, \ and\ \bibinfo {author} {\bibfnamefont {D.~J.}\
  \bibnamefont {Wineland}},\ }\href {\doibase 10.1038/nature02608} {\bibfield
  {journal} {\bibinfo  {journal} {Nature}\ }\textbf {\bibinfo {volume} {429}},\
  \bibinfo {pages} {737} (\bibinfo {year} {2004})}\BibitemShut {NoStop}%
\bibitem [{\citenamefont {Braunstein}\ and\ \citenamefont {van
  Loock}(2005)}]{Braunstein.RevModPhys.77.513(2005)}%
  \BibitemOpen
  \bibfield  {author} {\bibinfo {author} {\bibfnamefont {S.~L.}\ \bibnamefont
  {Braunstein}}\ and\ \bibinfo {author} {\bibfnamefont {P.}~\bibnamefont {van
  Loock}},\ }\href {\doibase 10.1103/RevModPhys.77.513} {\bibfield  {journal}
  {\bibinfo  {journal} {Rev. Mod. Phys.}\ }\textbf {\bibinfo {volume} {77}},\
  \bibinfo {pages} {513} (\bibinfo {year} {2005})}\BibitemShut {NoStop}%
\bibitem [{\citenamefont {Furusawa}\ \emph {et~al.}(1998)\citenamefont
  {Furusawa}, \citenamefont {S{\o}rensen}, \citenamefont {Braunstein},
  \citenamefont {Fuchs}, \citenamefont {Kimble},\ and\ \citenamefont
  {Polzik}}]{Furusawa.Science.282.706(1998)}%
  \BibitemOpen
  \bibfield  {author} {\bibinfo {author} {\bibfnamefont {A.}~\bibnamefont
  {Furusawa}}, \bibinfo {author} {\bibfnamefont {J.~L.}\ \bibnamefont
  {S{\o}rensen}}, \bibinfo {author} {\bibfnamefont {S.~L.}\ \bibnamefont
  {Braunstein}}, \bibinfo {author} {\bibfnamefont {C.~A.}\ \bibnamefont
  {Fuchs}}, \bibinfo {author} {\bibfnamefont {H.~J.}\ \bibnamefont {Kimble}}, \
  and\ \bibinfo {author} {\bibfnamefont {E.~S.}\ \bibnamefont {Polzik}},\
  }\href {\doibase 10.1126/science.282.5389.706} {\bibfield  {journal}
  {\bibinfo  {journal} {Science}\ }\textbf {\bibinfo {volume} {282}},\ \bibinfo
  {pages} {706} (\bibinfo {year} {1998})}\BibitemShut {NoStop}%
\bibitem [{\citenamefont {Sherson}\ \emph {et~al.}(2006)\citenamefont
  {Sherson}, \citenamefont {Krauter}, \citenamefont {Olsson}, \citenamefont
  {Julsgaard}, \citenamefont {Hammerer}, \citenamefont {Cirac},\ and\
  \citenamefont {Polzik}}]{Sherson.Nature.443.557(2006)}%
  \BibitemOpen
  \bibfield  {author} {\bibinfo {author} {\bibfnamefont {J.~F.}\ \bibnamefont
  {Sherson}}, \bibinfo {author} {\bibfnamefont {H.}~\bibnamefont {Krauter}},
  \bibinfo {author} {\bibfnamefont {R.~K.}\ \bibnamefont {Olsson}}, \bibinfo
  {author} {\bibfnamefont {B.}~\bibnamefont {Julsgaard}}, \bibinfo {author}
  {\bibfnamefont {K.}~\bibnamefont {Hammerer}}, \bibinfo {author}
  {\bibfnamefont {I.}~\bibnamefont {Cirac}}, \ and\ \bibinfo {author}
  {\bibfnamefont {E.~S.}\ \bibnamefont {Polzik}},\ }\href {\doibase
  10.1038/nature05136} {\bibfield  {journal} {\bibinfo  {journal} {Nature}\
  }\textbf {\bibinfo {volume} {443}},\ \bibinfo {pages} {557} (\bibinfo {year}
  {2006})}\BibitemShut {NoStop}%
\bibitem [{\citenamefont {Krauter}\ \emph {et~al.}(2013)\citenamefont
  {Krauter}, \citenamefont {Salart}, \citenamefont {Muschik}, \citenamefont
  {Petersen}, \citenamefont {Shen}, \citenamefont {Fernholz},\ and\
  \citenamefont {Polzik}}]{Krauter.NaturePhys.9.400(2013)}%
  \BibitemOpen
  \bibfield  {author} {\bibinfo {author} {\bibfnamefont {H.}~\bibnamefont
  {Krauter}}, \bibinfo {author} {\bibfnamefont {D.}~\bibnamefont {Salart}},
  \bibinfo {author} {\bibfnamefont {C.~A.}\ \bibnamefont {Muschik}}, \bibinfo
  {author} {\bibfnamefont {J.~M.}\ \bibnamefont {Petersen}}, \bibinfo {author}
  {\bibfnamefont {H.}~\bibnamefont {Shen}}, \bibinfo {author} {\bibfnamefont
  {T.}~\bibnamefont {Fernholz}}, \ and\ \bibinfo {author} {\bibfnamefont
  {E.~S.}\ \bibnamefont {Polzik}},\ }\href {\doibase 10.1038/NPHYS2631}
  {\bibfield  {journal} {\bibinfo  {journal} {Nature Phys.}\ }\textbf {\bibinfo
  {volume} {9}},\ \bibinfo {pages} {400} (\bibinfo {year} {2013})}\BibitemShut
  {NoStop}%
\bibitem [{\citenamefont {Julsgaard}\ \emph {et~al.}(2004)\citenamefont
  {Julsgaard}, \citenamefont {Sherson}, \citenamefont {Cirac}, \citenamefont
  {Fiur\`{a}\v{s}ek},\ and\ \citenamefont
  {Polzik}}]{Julsgaard.Nature.432.482(2004)}%
  \BibitemOpen
  \bibfield  {author} {\bibinfo {author} {\bibfnamefont {B.}~\bibnamefont
  {Julsgaard}}, \bibinfo {author} {\bibfnamefont {J.}~\bibnamefont {Sherson}},
  \bibinfo {author} {\bibfnamefont {J.~I.}\ \bibnamefont {Cirac}}, \bibinfo
  {author} {\bibfnamefont {J.}~\bibnamefont {Fiur\`{a}\v{s}ek}}, \ and\
  \bibinfo {author} {\bibfnamefont {E.~S.}\ \bibnamefont {Polzik}},\ }\href
  {\doibase 10.1038/nature03064} {\bibfield  {journal} {\bibinfo  {journal}
  {Nature}\ }\textbf {\bibinfo {volume} {432}},\ \bibinfo {pages} {482}
  (\bibinfo {year} {2004})}\BibitemShut {NoStop}%
\bibitem [{\citenamefont {Honda}\ \emph {et~al.}(2008)\citenamefont {Honda},
  \citenamefont {Akamatsu}, \citenamefont {Arikawa}, \citenamefont {Yokoi},
  \citenamefont {Akiba}, \citenamefont {Nagatsuka}, \citenamefont {Tanimura},
  \citenamefont {Furusawa},\ and\ \citenamefont
  {Kozuma}}]{Honda.PhysRevLett.100.093601(2008)}%
  \BibitemOpen
  \bibfield  {author} {\bibinfo {author} {\bibfnamefont {K.}~\bibnamefont
  {Honda}}, \bibinfo {author} {\bibfnamefont {D.}~\bibnamefont {Akamatsu}},
  \bibinfo {author} {\bibfnamefont {M.}~\bibnamefont {Arikawa}}, \bibinfo
  {author} {\bibfnamefont {Y.}~\bibnamefont {Yokoi}}, \bibinfo {author}
  {\bibfnamefont {K.}~\bibnamefont {Akiba}}, \bibinfo {author} {\bibfnamefont
  {S.}~\bibnamefont {Nagatsuka}}, \bibinfo {author} {\bibfnamefont
  {T.}~\bibnamefont {Tanimura}}, \bibinfo {author} {\bibfnamefont
  {A.}~\bibnamefont {Furusawa}}, \ and\ \bibinfo {author} {\bibfnamefont
  {M.}~\bibnamefont {Kozuma}},\ }\href {\doibase
  10.1103/PhysRevLett.100.093601} {\bibfield  {journal} {\bibinfo  {journal}
  {Phys. Rev. Lett.}\ }\textbf {\bibinfo {volume} {100}},\ \bibinfo {pages}
  {093601} (\bibinfo {year} {2008})}\BibitemShut {NoStop}%
\bibitem [{\citenamefont {Appel}\ \emph {et~al.}(2008)\citenamefont {Appel},
  \citenamefont {Figueroa}, \citenamefont {Korystov}, \citenamefont {Lobino},\
  and\ \citenamefont {Lvovsky}}]{Appel.PhysRevLett.100.093602(2008)}%
  \BibitemOpen
  \bibfield  {author} {\bibinfo {author} {\bibfnamefont {J.}~\bibnamefont
  {Appel}}, \bibinfo {author} {\bibfnamefont {E.}~\bibnamefont {Figueroa}},
  \bibinfo {author} {\bibfnamefont {D.}~\bibnamefont {Korystov}}, \bibinfo
  {author} {\bibfnamefont {M.}~\bibnamefont {Lobino}}, \ and\ \bibinfo {author}
  {\bibfnamefont {A.~I.}\ \bibnamefont {Lvovsky}},\ }\href {\doibase
  10.1103/PhysRevLett.100.093602} {\bibfield  {journal} {\bibinfo  {journal}
  {Phys. Rev. Lett.}\ }\textbf {\bibinfo {volume} {100}},\ \bibinfo {pages}
  {093602} (\bibinfo {year} {2008})}\BibitemShut {NoStop}%
\bibitem [{\citenamefont {{de Riedmatten}}\ \emph {et~al.}(2008)\citenamefont
  {{de Riedmatten}}, \citenamefont {Afzelius}, \citenamefont {Staudt},
  \citenamefont {Simon},\ and\ \citenamefont
  {Gisin}}]{deRiedmatten.Nature.456.773(2008)}%
  \BibitemOpen
  \bibfield  {author} {\bibinfo {author} {\bibfnamefont {H.}~\bibnamefont {{de
  Riedmatten}}}, \bibinfo {author} {\bibfnamefont {M.}~\bibnamefont
  {Afzelius}}, \bibinfo {author} {\bibfnamefont {M.~U.}\ \bibnamefont
  {Staudt}}, \bibinfo {author} {\bibfnamefont {C.}~\bibnamefont {Simon}}, \
  and\ \bibinfo {author} {\bibfnamefont {N.}~\bibnamefont {Gisin}},\ }\href
  {\doibase 10.1038/nature07607} {\bibfield  {journal} {\bibinfo  {journal}
  {Nature}\ }\textbf {\bibinfo {volume} {456}},\ \bibinfo {pages} {773}
  (\bibinfo {year} {2008})}\BibitemShut {NoStop}%
\bibitem [{\citenamefont {Hedges}\ \emph {et~al.}(2010)\citenamefont {Hedges},
  \citenamefont {Longdell}, \citenamefont {Li},\ and\ \citenamefont
  {Sellars}}]{Hedges.Nature.465.1052(2010)}%
  \BibitemOpen
  \bibfield  {author} {\bibinfo {author} {\bibfnamefont {M.~P.}\ \bibnamefont
  {Hedges}}, \bibinfo {author} {\bibfnamefont {J.~J.}\ \bibnamefont
  {Longdell}}, \bibinfo {author} {\bibfnamefont {Y.}~\bibnamefont {Li}}, \ and\
  \bibinfo {author} {\bibfnamefont {M.~J.}\ \bibnamefont {Sellars}},\ }\href
  {\doibase 10.1038/nature09081} {\bibfield  {journal} {\bibinfo  {journal}
  {Nature}\ }\textbf {\bibinfo {volume} {465}},\ \bibinfo {pages} {1052}
  (\bibinfo {year} {2010})}\BibitemShut {NoStop}%
\bibitem [{\citenamefont {Hammerer}\ \emph {et~al.}(2010)\citenamefont
  {Hammerer}, \citenamefont {S{\o}rensen},\ and\ \citenamefont
  {Polzik}}]{Hammerer.RevModPhys.82.1041(2010)}%
  \BibitemOpen
  \bibfield  {author} {\bibinfo {author} {\bibfnamefont {K.}~\bibnamefont
  {Hammerer}}, \bibinfo {author} {\bibfnamefont {A.~S.}\ \bibnamefont
  {S{\o}rensen}}, \ and\ \bibinfo {author} {\bibfnamefont {E.~S.}\ \bibnamefont
  {Polzik}},\ }\href {\doibase 10.1103/RevModPhys.82.1041} {\bibfield
  {journal} {\bibinfo  {journal} {Rev. Mod. Phys.}\ }\textbf {\bibinfo {volume}
  {82}},\ \bibinfo {pages} {1041} (\bibinfo {year} {2010})}\BibitemShut
  {NoStop}%
\bibitem [{\citenamefont {Tittel}\ \emph {et~al.}(2010)\citenamefont {Tittel},
  \citenamefont {Afzelius}, \citenamefont {Chaneli\`ere}, \citenamefont {Cone},
  \citenamefont {Kr\"oll}, \citenamefont {Moiseev},\ and\ \citenamefont
  {Sellars}}]{Tittel.LaserPhotonRev.4.244(2010)}%
  \BibitemOpen
  \bibfield  {author} {\bibinfo {author} {\bibfnamefont {W.}~\bibnamefont
  {Tittel}}, \bibinfo {author} {\bibfnamefont {M.}~\bibnamefont {Afzelius}},
  \bibinfo {author} {\bibfnamefont {T.}~\bibnamefont {Chaneli\`ere}}, \bibinfo
  {author} {\bibfnamefont {R.~L.}\ \bibnamefont {Cone}}, \bibinfo {author}
  {\bibfnamefont {S.}~\bibnamefont {Kr\"oll}}, \bibinfo {author} {\bibfnamefont
  {S.~A.}\ \bibnamefont {Moiseev}}, \ and\ \bibinfo {author} {\bibfnamefont
  {M.}~\bibnamefont {Sellars}},\ }\href {\doibase 10.1002/lpor.200810056}
  {\bibfield  {journal} {\bibinfo  {journal} {Laser \& Photon. Rev.}\ }\textbf
  {\bibinfo {volume} {4}},\ \bibinfo {pages} {244} (\bibinfo {year}
  {2010})}\BibitemShut {NoStop}%
\bibitem [{\citenamefont {Kleckner}\ and\ \citenamefont
  {Bouwmeester}(2006)}]{Kleckner.Nature.444.75(2006)}%
  \BibitemOpen
  \bibfield  {author} {\bibinfo {author} {\bibfnamefont {D.}~\bibnamefont
  {Kleckner}}\ and\ \bibinfo {author} {\bibfnamefont {D.}~\bibnamefont
  {Bouwmeester}},\ }\href {\doibase 10.1038/nature05231} {\bibfield  {journal}
  {\bibinfo  {journal} {Nature}\ }\textbf {\bibinfo {volume} {444}},\ \bibinfo
  {pages} {75} (\bibinfo {year} {2006})}\BibitemShut {NoStop}%
\bibitem [{\citenamefont {Thompson}\ \emph {et~al.}(2008)\citenamefont
  {Thompson}, \citenamefont {Zwickl}, \citenamefont {Jayich}, \citenamefont
  {Marquardt}, \citenamefont {Girvin},\ and\ \citenamefont
  {Harris}}]{Thompson.Nature.452.72(2008)}%
  \BibitemOpen
  \bibfield  {author} {\bibinfo {author} {\bibfnamefont {J.~D.}\ \bibnamefont
  {Thompson}}, \bibinfo {author} {\bibfnamefont {B.~M.}\ \bibnamefont
  {Zwickl}}, \bibinfo {author} {\bibfnamefont {A.~M.}\ \bibnamefont {Jayich}},
  \bibinfo {author} {\bibfnamefont {F.}~\bibnamefont {Marquardt}}, \bibinfo
  {author} {\bibfnamefont {S.~M.}\ \bibnamefont {Girvin}}, \ and\ \bibinfo
  {author} {\bibfnamefont {J.~G.~E.}\ \bibnamefont {Harris}},\ }\href {\doibase
  10.1038/nature06715} {\bibfield  {journal} {\bibinfo  {journal} {Nature}\
  }\textbf {\bibinfo {volume} {452}},\ \bibinfo {pages} {72} (\bibinfo {year}
  {2008})}\BibitemShut {NoStop}%
\bibitem [{\citenamefont {Kippenberg}\ \emph {et~al.}(2005)\citenamefont
  {Kippenberg}, \citenamefont {Rokhsari}, \citenamefont {Carmon}, \citenamefont
  {Scherer},\ and\ \citenamefont
  {Vahala}}]{Kippenberg.PhysRevLett.95.033901(2005)}%
  \BibitemOpen
  \bibfield  {author} {\bibinfo {author} {\bibfnamefont {T.~J.}\ \bibnamefont
  {Kippenberg}}, \bibinfo {author} {\bibfnamefont {H.}~\bibnamefont
  {Rokhsari}}, \bibinfo {author} {\bibfnamefont {T.}~\bibnamefont {Carmon}},
  \bibinfo {author} {\bibfnamefont {A.}~\bibnamefont {Scherer}}, \ and\
  \bibinfo {author} {\bibfnamefont {K.~J.}\ \bibnamefont {Vahala}},\ }\href
  {\doibase 10.1103/PhysRevLett.95.033901} {\bibfield  {journal} {\bibinfo
  {journal} {Phys. Rev. Lett.}\ }\textbf {\bibinfo {volume} {95}},\ \bibinfo
  {pages} {033901} (\bibinfo {year} {2005})}\BibitemShut {NoStop}%
\bibitem [{\citenamefont {Regal}\ \emph {et~al.}(2008)\citenamefont {Regal},
  \citenamefont {Teufel},\ and\ \citenamefont
  {Lehnert}}]{Regal.NaturePhys.4.555(2008)}%
  \BibitemOpen
  \bibfield  {author} {\bibinfo {author} {\bibfnamefont {C.~A.}\ \bibnamefont
  {Regal}}, \bibinfo {author} {\bibfnamefont {J.~D.}\ \bibnamefont {Teufel}}, \
  and\ \bibinfo {author} {\bibfnamefont {K.~W.}\ \bibnamefont {Lehnert}},\
  }\href {\doibase 10.1038/nphys974} {\bibfield  {journal} {\bibinfo  {journal}
  {Nature Phys.}\ }\textbf {\bibinfo {volume} {4}},\ \bibinfo {pages} {555}
  (\bibinfo {year} {2008})}\BibitemShut {NoStop}%
\bibitem [{\citenamefont {Hunger}\ \emph {et~al.}(2010)\citenamefont {Hunger},
  \citenamefont {Camerer}, \citenamefont {H\"ansch}, \citenamefont {K\"onig},
  \citenamefont {Kotthaus}, \citenamefont {Reichel},\ and\ \citenamefont
  {Treutlein}}]{Hunger.PhysRevLett.104.143002(2010)}%
  \BibitemOpen
  \bibfield  {author} {\bibinfo {author} {\bibfnamefont {D.}~\bibnamefont
  {Hunger}}, \bibinfo {author} {\bibfnamefont {S.}~\bibnamefont {Camerer}},
  \bibinfo {author} {\bibfnamefont {T.~W.}\ \bibnamefont {H\"ansch}}, \bibinfo
  {author} {\bibfnamefont {D.}~\bibnamefont {K\"onig}}, \bibinfo {author}
  {\bibfnamefont {J.~P.}\ \bibnamefont {Kotthaus}}, \bibinfo {author}
  {\bibfnamefont {J.}~\bibnamefont {Reichel}}, \ and\ \bibinfo {author}
  {\bibfnamefont {P.}~\bibnamefont {Treutlein}},\ }\href {\doibase
  10.1103/PhysRevLett.104.143002} {\bibfield  {journal} {\bibinfo  {journal}
  {Phys. Rev. Lett.}\ }\textbf {\bibinfo {volume} {104}},\ \bibinfo {pages}
  {143002} (\bibinfo {year} {2010})}\BibitemShut {NoStop}%
\bibitem [{\citenamefont {Chou}\ \emph {et~al.}(2005)\citenamefont {Chou},
  \citenamefont {{de Riedmatten}}, \citenamefont {Felinto}, \citenamefont
  {Polyakov}, \citenamefont {{van Enk}},\ and\ \citenamefont
  {Kimble}}]{Chou.Nature.438.828(2005)}%
  \BibitemOpen
  \bibfield  {author} {\bibinfo {author} {\bibfnamefont {C.~W.}\ \bibnamefont
  {Chou}}, \bibinfo {author} {\bibfnamefont {H.}~\bibnamefont {{de
  Riedmatten}}}, \bibinfo {author} {\bibfnamefont {D.}~\bibnamefont {Felinto}},
  \bibinfo {author} {\bibfnamefont {S.~V.}\ \bibnamefont {Polyakov}}, \bibinfo
  {author} {\bibfnamefont {S.~J.}\ \bibnamefont {{van Enk}}}, \ and\ \bibinfo
  {author} {\bibfnamefont {H.~J.}\ \bibnamefont {Kimble}},\ }\href {\doibase
  10.1038/nature04353} {\bibfield  {journal} {\bibinfo  {journal} {Nature}\
  }\textbf {\bibinfo {volume} {438}},\ \bibinfo {pages} {828} (\bibinfo {year}
  {2005})}\BibitemShut {NoStop}%
\bibitem [{\citenamefont {Chaneli\`ere}\ \emph {et~al.}(2005)\citenamefont
  {Chaneli\`ere}, \citenamefont {Matsukevich}, \citenamefont {Jenkins},
  \citenamefont {Lan}, \citenamefont {Kennedy},\ and\ \citenamefont
  {Kuzmich}}]{Chaneliere.Nature.438.833(2005)}%
  \BibitemOpen
  \bibfield  {author} {\bibinfo {author} {\bibfnamefont {T.}~\bibnamefont
  {Chaneli\`ere}}, \bibinfo {author} {\bibfnamefont {D.~N.}\ \bibnamefont
  {Matsukevich}}, \bibinfo {author} {\bibfnamefont {S.~D.}\ \bibnamefont
  {Jenkins}}, \bibinfo {author} {\bibfnamefont {S.}~\bibnamefont {Lan}},
  \bibinfo {author} {\bibfnamefont {T.~A.~B.}\ \bibnamefont {Kennedy}}, \ and\
  \bibinfo {author} {\bibfnamefont {A.}~\bibnamefont {Kuzmich}},\ }\href
  {\doibase 10.1038/nature04315} {\bibfield  {journal} {\bibinfo  {journal}
  {Nature}\ }\textbf {\bibinfo {volume} {438}},\ \bibinfo {pages} {833}
  (\bibinfo {year} {2005})}\BibitemShut {NoStop}%
\bibitem [{\citenamefont {Eisaman}\ \emph {et~al.}(2005)\citenamefont
  {Eisaman}, \citenamefont {Andr\'e}, \citenamefont {Massou}, \citenamefont
  {Fleischhauer}, \citenamefont {Zibrov},\ and\ \citenamefont
  {Lukin}}]{Eisaman.Nature.438.837(2005)}%
  \BibitemOpen
  \bibfield  {author} {\bibinfo {author} {\bibfnamefont {M.~D.}\ \bibnamefont
  {Eisaman}}, \bibinfo {author} {\bibfnamefont {A.}~\bibnamefont {Andr\'e}},
  \bibinfo {author} {\bibfnamefont {F.}~\bibnamefont {Massou}}, \bibinfo
  {author} {\bibfnamefont {M.}~\bibnamefont {Fleischhauer}}, \bibinfo {author}
  {\bibfnamefont {A.~S.}\ \bibnamefont {Zibrov}}, \ and\ \bibinfo {author}
  {\bibfnamefont {M.~D.}\ \bibnamefont {Lukin}},\ }\href {\doibase
  10.1038/nature04327} {\bibfield  {journal} {\bibinfo  {journal} {Nature}\
  }\textbf {\bibinfo {volume} {438}},\ \bibinfo {pages} {837} (\bibinfo {year}
  {2005})}\BibitemShut {NoStop}%
\bibitem [{\citenamefont {Kubo}\ \emph {et~al.}(2011)\citenamefont {Kubo},
  \citenamefont {Grezes}, \citenamefont {Dewes}, \citenamefont {Umeda},
  \citenamefont {Isoya}, \citenamefont {Sumiya}, \citenamefont {Morishita},
  \citenamefont {Abe}, \citenamefont {Onoda}, \citenamefont {Ohshima},
  \citenamefont {Jacques}, \citenamefont {Dr\'eau}, \citenamefont {Roch},
  \citenamefont {Diniz}, \citenamefont {Auffeves}, \citenamefont {Vion},
  \citenamefont {Esteve},\ and\ \citenamefont
  {Bertet}}]{Kubo.PhysRevLett.107.220501(2011)}%
  \BibitemOpen
  \bibfield  {author} {\bibinfo {author} {\bibfnamefont {Y.}~\bibnamefont
  {Kubo}}, \bibinfo {author} {\bibfnamefont {C.}~\bibnamefont {Grezes}},
  \bibinfo {author} {\bibfnamefont {A.}~\bibnamefont {Dewes}}, \bibinfo
  {author} {\bibfnamefont {T.}~\bibnamefont {Umeda}}, \bibinfo {author}
  {\bibfnamefont {J.}~\bibnamefont {Isoya}}, \bibinfo {author} {\bibfnamefont
  {H.}~\bibnamefont {Sumiya}}, \bibinfo {author} {\bibfnamefont
  {N.}~\bibnamefont {Morishita}}, \bibinfo {author} {\bibfnamefont
  {H.}~\bibnamefont {Abe}}, \bibinfo {author} {\bibfnamefont {S.}~\bibnamefont
  {Onoda}}, \bibinfo {author} {\bibfnamefont {T.}~\bibnamefont {Ohshima}},
  \bibinfo {author} {\bibfnamefont {V.}~\bibnamefont {Jacques}}, \bibinfo
  {author} {\bibfnamefont {A.}~\bibnamefont {Dr\'eau}}, \bibinfo {author}
  {\bibfnamefont {J.-F.}\ \bibnamefont {Roch}}, \bibinfo {author}
  {\bibfnamefont {I.}~\bibnamefont {Diniz}}, \bibinfo {author} {\bibfnamefont
  {A.}~\bibnamefont {Auffeves}}, \bibinfo {author} {\bibfnamefont
  {D.}~\bibnamefont {Vion}}, \bibinfo {author} {\bibfnamefont {D.}~\bibnamefont
  {Esteve}}, \ and\ \bibinfo {author} {\bibfnamefont {P.}~\bibnamefont
  {Bertet}},\ }\href {\doibase 10.1103/PhysRevLett.107.220501} {\bibfield
  {journal} {\bibinfo  {journal} {Phys. Rev. Lett.}\ }\textbf {\bibinfo
  {volume} {107}},\ \bibinfo {pages} {220501} (\bibinfo {year}
  {2011})}\BibitemShut {NoStop}%
\bibitem [{\citenamefont {Clausen}\ \emph {et~al.}(2012)\citenamefont
  {Clausen}, \citenamefont {Bussi\`eres}, \citenamefont {Afzelius},\ and\
  \citenamefont {Gisin}}]{Clausen.PhysRevLett.108.190503(2012)}%
  \BibitemOpen
  \bibfield  {author} {\bibinfo {author} {\bibfnamefont {C.}~\bibnamefont
  {Clausen}}, \bibinfo {author} {\bibfnamefont {F.}~\bibnamefont
  {Bussi\`eres}}, \bibinfo {author} {\bibfnamefont {M.}~\bibnamefont
  {Afzelius}}, \ and\ \bibinfo {author} {\bibfnamefont {N.}~\bibnamefont
  {Gisin}},\ }\href {\doibase 10.1103/PhysRevLett.108.190503} {\bibfield
  {journal} {\bibinfo  {journal} {Phys. Rev. Lett.}\ }\textbf {\bibinfo
  {volume} {108}},\ \bibinfo {pages} {190503} (\bibinfo {year}
  {2012})}\BibitemShut {NoStop}%
\bibitem [{\citenamefont {G\"undo\u{g}an}\ \emph {et~al.}(2012)\citenamefont
  {G\"undo\u{g}an}, \citenamefont {Ledingham}, \citenamefont {Almasi},
  \citenamefont {Cristiani},\ and\ \citenamefont {{de
  Riedmatten}}}]{Gundogan.PhysRevLett.108.190504(2012)}%
  \BibitemOpen
  \bibfield  {author} {\bibinfo {author} {\bibfnamefont {M.}~\bibnamefont
  {G\"undo\u{g}an}}, \bibinfo {author} {\bibfnamefont {P.~M.}\ \bibnamefont
  {Ledingham}}, \bibinfo {author} {\bibfnamefont {A.}~\bibnamefont {Almasi}},
  \bibinfo {author} {\bibfnamefont {M.}~\bibnamefont {Cristiani}}, \ and\
  \bibinfo {author} {\bibfnamefont {H.}~\bibnamefont {{de Riedmatten}}},\
  }\href {\doibase 10.1103/PhysRevLett.108.190504} {\bibfield  {journal}
  {\bibinfo  {journal} {Phys. Rev. Lett.}\ }\textbf {\bibinfo {volume} {108}},\
  \bibinfo {pages} {190504} (\bibinfo {year} {2012})}\BibitemShut {NoStop}%
\bibitem [{\citenamefont {Xiang}\ \emph {et~al.}(2013)\citenamefont {Xiang},
  \citenamefont {Ashhab}, \citenamefont {You},\ and\ \citenamefont
  {Nori}}]{Xiang.RevModPhys.85.623(2013)}%
  \BibitemOpen
  \bibfield  {author} {\bibinfo {author} {\bibfnamefont {Z.-L.}\ \bibnamefont
  {Xiang}}, \bibinfo {author} {\bibfnamefont {S.}~\bibnamefont {Ashhab}},
  \bibinfo {author} {\bibfnamefont {J.~Q.}\ \bibnamefont {You}}, \ and\
  \bibinfo {author} {\bibfnamefont {F.}~\bibnamefont {Nori}},\ }\href {\doibase
  10.1103/RevModPhys.85.623} {\bibfield  {journal} {\bibinfo  {journal} {Rev.
  Mod. Phys.}\ }\textbf {\bibinfo {volume} {85}},\ \bibinfo {pages} {623}
  (\bibinfo {year} {2013})}\BibitemShut {NoStop}%
\bibitem [{\citenamefont {Rabl}\ \emph {et~al.}(2006)\citenamefont {Rabl},
  \citenamefont {DeMille}, \citenamefont {Doyle}, \citenamefont {Lukin},
  \citenamefont {Schoelkopf},\ and\ \citenamefont
  {Zoller}}]{Rabl.PhysRevLett.97.033003(2006)}%
  \BibitemOpen
  \bibfield  {author} {\bibinfo {author} {\bibfnamefont {P.}~\bibnamefont
  {Rabl}}, \bibinfo {author} {\bibfnamefont {D.}~\bibnamefont {DeMille}},
  \bibinfo {author} {\bibfnamefont {J.~M.}\ \bibnamefont {Doyle}}, \bibinfo
  {author} {\bibfnamefont {M.~D.}\ \bibnamefont {Lukin}}, \bibinfo {author}
  {\bibfnamefont {R.~J.}\ \bibnamefont {Schoelkopf}}, \ and\ \bibinfo {author}
  {\bibfnamefont {P.}~\bibnamefont {Zoller}},\ }\href {\doibase
  10.1103/PhysRevLett.97.033003} {\bibfield  {journal} {\bibinfo  {journal}
  {Phys. Rev. Lett.}\ }\textbf {\bibinfo {volume} {97}},\ \bibinfo {pages}
  {033003} (\bibinfo {year} {2006})}\BibitemShut {NoStop}%
\bibitem [{\citenamefont {Petrosyan}\ \emph {et~al.}(2009)\citenamefont
  {Petrosyan}, \citenamefont {Bensky}, \citenamefont {Kurizki}, \citenamefont
  {Mazets}, \citenamefont {Majer},\ and\ \citenamefont
  {Schmiedmayer}}]{Petrosyan.PhysRevA.79.040304R(2009)}%
  \BibitemOpen
  \bibfield  {author} {\bibinfo {author} {\bibfnamefont {D.}~\bibnamefont
  {Petrosyan}}, \bibinfo {author} {\bibfnamefont {G.}~\bibnamefont {Bensky}},
  \bibinfo {author} {\bibfnamefont {G.}~\bibnamefont {Kurizki}}, \bibinfo
  {author} {\bibfnamefont {I.}~\bibnamefont {Mazets}}, \bibinfo {author}
  {\bibfnamefont {J.}~\bibnamefont {Majer}}, \ and\ \bibinfo {author}
  {\bibfnamefont {J.}~\bibnamefont {Schmiedmayer}},\ }\href {\doibase
  10.1103/PhysRevA.79.040304} {\bibfield  {journal} {\bibinfo  {journal} {Phys.
  Rev. A}\ }\textbf {\bibinfo {volume} {79}},\ \bibinfo {pages} {040304(R)}
  (\bibinfo {year} {2009})}\BibitemShut {NoStop}%
\bibitem [{\citenamefont {Wesenberg}\ \emph {et~al.}(2009)\citenamefont
  {Wesenberg}, \citenamefont {Ardavan}, \citenamefont {Briggs}, \citenamefont
  {Morton}, \citenamefont {Schoelkopf}, \citenamefont {Schuster},\ and\
  \citenamefont {M{\o}lmer}}]{Wesenberg.PhysRevLett.103.070502(2009)}%
  \BibitemOpen
  \bibfield  {author} {\bibinfo {author} {\bibfnamefont {J.~H.}\ \bibnamefont
  {Wesenberg}}, \bibinfo {author} {\bibfnamefont {A.}~\bibnamefont {Ardavan}},
  \bibinfo {author} {\bibfnamefont {G.~A.~D.}\ \bibnamefont {Briggs}}, \bibinfo
  {author} {\bibfnamefont {J.~J.~L.}\ \bibnamefont {Morton}}, \bibinfo {author}
  {\bibfnamefont {R.~J.}\ \bibnamefont {Schoelkopf}}, \bibinfo {author}
  {\bibfnamefont {D.~I.}\ \bibnamefont {Schuster}}, \ and\ \bibinfo {author}
  {\bibfnamefont {K.}~\bibnamefont {M{\o}lmer}},\ }\href {\doibase
  10.1103/PhysRevLett.103.070502} {\bibfield  {journal} {\bibinfo  {journal}
  {Phys. Rev. Lett.}\ }\textbf {\bibinfo {volume} {103}},\ \bibinfo {pages}
  {070502} (\bibinfo {year} {2009})}\BibitemShut {NoStop}%
\bibitem [{\citenamefont {Marcos}\ \emph {et~al.}(2010)\citenamefont {Marcos},
  \citenamefont {Wubs}, \citenamefont {Taylor}, \citenamefont {Aguado},
  \citenamefont {Lukin},\ and\ \citenamefont
  {S{\o}rensen}}]{Marcos.PhysRevLett.105.210501(2010)}%
  \BibitemOpen
  \bibfield  {author} {\bibinfo {author} {\bibfnamefont {D.}~\bibnamefont
  {Marcos}}, \bibinfo {author} {\bibfnamefont {M.}~\bibnamefont {Wubs}},
  \bibinfo {author} {\bibfnamefont {J.~M.}\ \bibnamefont {Taylor}}, \bibinfo
  {author} {\bibfnamefont {R.}~\bibnamefont {Aguado}}, \bibinfo {author}
  {\bibfnamefont {M.~D.}\ \bibnamefont {Lukin}}, \ and\ \bibinfo {author}
  {\bibfnamefont {A.~S.}\ \bibnamefont {S{\o}rensen}},\ }\href {\doibase
  10.1103/PhysRevLett.105.210501} {\bibfield  {journal} {\bibinfo  {journal}
  {Phys. Rev. Lett.}\ }\textbf {\bibinfo {volume} {105}},\ \bibinfo {pages}
  {210501} (\bibinfo {year} {2010})}\BibitemShut {NoStop}%
\bibitem [{\citenamefont {{O'Connell}}\ \emph {et~al.}(2010)\citenamefont
  {{O'Connell}}, \citenamefont {Hofheinz}, \citenamefont {Ansmann},
  \citenamefont {Bialczak}, \citenamefont {Lenander}, \citenamefont {Lucero},
  \citenamefont {Neeley}, \citenamefont {Sank}, \citenamefont {Wang},
  \citenamefont {Weides}, \citenamefont {Wenner}, \citenamefont {Martinis},\
  and\ \citenamefont {Cleland}}]{OConnell.Nature.464.697(2010)}%
  \BibitemOpen
  \bibfield  {author} {\bibinfo {author} {\bibfnamefont {A.~D.}\ \bibnamefont
  {{O'Connell}}}, \bibinfo {author} {\bibfnamefont {M.}~\bibnamefont
  {Hofheinz}}, \bibinfo {author} {\bibfnamefont {M.}~\bibnamefont {Ansmann}},
  \bibinfo {author} {\bibfnamefont {R.~C.}\ \bibnamefont {Bialczak}}, \bibinfo
  {author} {\bibfnamefont {M.}~\bibnamefont {Lenander}}, \bibinfo {author}
  {\bibfnamefont {E.}~\bibnamefont {Lucero}}, \bibinfo {author} {\bibfnamefont
  {M.}~\bibnamefont {Neeley}}, \bibinfo {author} {\bibfnamefont
  {D.}~\bibnamefont {Sank}}, \bibinfo {author} {\bibfnamefont {H.}~\bibnamefont
  {Wang}}, \bibinfo {author} {\bibfnamefont {M.}~\bibnamefont {Weides}},
  \bibinfo {author} {\bibfnamefont {J.}~\bibnamefont {Wenner}}, \bibinfo
  {author} {\bibfnamefont {J.~M.}\ \bibnamefont {Martinis}}, \ and\ \bibinfo
  {author} {\bibfnamefont {A.~N.}\ \bibnamefont {Cleland}},\ }\href {\doibase
  10.1038/nature08967} {\bibfield  {journal} {\bibinfo  {journal} {Nature}\
  }\textbf {\bibinfo {volume} {464}},\ \bibinfo {pages} {697} (\bibinfo {year}
  {2010})}\BibitemShut {NoStop}%
\bibitem [{\citenamefont {Arcizet}\ \emph {et~al.}(2011)\citenamefont
  {Arcizet}, \citenamefont {Jacques}, \citenamefont {Siria}, \citenamefont
  {Poncharal}, \citenamefont {Vincent},\ and\ \citenamefont
  {Seidelin}}]{Arcizet.NaturePhys.7.879(2011)}%
  \BibitemOpen
  \bibfield  {author} {\bibinfo {author} {\bibfnamefont {O.}~\bibnamefont
  {Arcizet}}, \bibinfo {author} {\bibfnamefont {V.}~\bibnamefont {Jacques}},
  \bibinfo {author} {\bibfnamefont {A.}~\bibnamefont {Siria}}, \bibinfo
  {author} {\bibfnamefont {P.}~\bibnamefont {Poncharal}}, \bibinfo {author}
  {\bibfnamefont {P.}~\bibnamefont {Vincent}}, \ and\ \bibinfo {author}
  {\bibfnamefont {S.}~\bibnamefont {Seidelin}},\ }\href {\doibase
  10.1038/NPHYS2070} {\bibfield  {journal} {\bibinfo  {journal} {Nature Phys.}\
  }\textbf {\bibinfo {volume} {7}},\ \bibinfo {pages} {879} (\bibinfo {year}
  {2011})}\BibitemShut {NoStop}%
\bibitem [{\citenamefont {\protect{J.-M.} Pirkkalainen}\ \emph
  {et~al.}(2013)\citenamefont {\protect{J.-M.} Pirkkalainen}, \citenamefont
  {Cho}, \citenamefont {Li}, \citenamefont {Paraoanu}, \citenamefont
  {Hakonen},\ and\ \citenamefont
  {Sillanp\"a\"a}}]{Pirkkalainen.Nature.494.211(2013)}%
  \BibitemOpen
  \bibfield  {author} {\bibinfo {author} {\bibnamefont {\protect{J.-M.}
  Pirkkalainen}}, \bibinfo {author} {\bibfnamefont {S.~U.}\ \bibnamefont
  {Cho}}, \bibinfo {author} {\bibfnamefont {J.}~\bibnamefont {Li}}, \bibinfo
  {author} {\bibfnamefont {G.~S.}\ \bibnamefont {Paraoanu}}, \bibinfo {author}
  {\bibfnamefont {P.~J.}\ \bibnamefont {Hakonen}}, \ and\ \bibinfo {author}
  {\bibfnamefont {M.~A.}\ \bibnamefont {Sillanp\"a\"a}},\ }\href {\doibase
  10.1038/nature11821} {\bibfield  {journal} {\bibinfo  {journal} {Nature}\
  }\textbf {\bibinfo {volume} {494}},\ \bibinfo {pages} {211} (\bibinfo {year}
  {2013})}\BibitemShut {NoStop}%
\bibitem [{\citenamefont {Wang}\ \emph {et~al.}(2013)\citenamefont {Wang},
  \citenamefont {Yu}, \citenamefont {Hu}, \citenamefont {Miranowicz},\ and\
  \citenamefont {Nori}}]{Wang.PhysRevA.88.022101(2013)}%
  \BibitemOpen
  \bibfield  {author} {\bibinfo {author} {\bibfnamefont {X.-B.}\ \bibnamefont
  {Wang}}, \bibinfo {author} {\bibfnamefont {Z.-W.}\ \bibnamefont {Yu}},
  \bibinfo {author} {\bibfnamefont {J.-Z.}\ \bibnamefont {Hu}}, \bibinfo
  {author} {\bibfnamefont {A.}~\bibnamefont {Miranowicz}}, \ and\ \bibinfo
  {author} {\bibfnamefont {F.}~\bibnamefont {Nori}},\ }\href {\doibase
  10.1103/PhysRevA.88.022101} {\bibfield  {journal} {\bibinfo  {journal} {Phys.
  Rev. A}\ }\textbf {\bibinfo {volume} {88}},\ \bibinfo {pages} {022101}
  (\bibinfo {year} {2013})}\BibitemShut {NoStop}%
\bibitem [{\citenamefont {Julsgaard}\ \emph {et~al.}(2013)\citenamefont
  {Julsgaard}, \citenamefont {Grezes}, \citenamefont {Bertet},\ and\
  \citenamefont {M{\o}lmer}}]{Julsgaard.PhysRevLett.110.250503(2013)}%
  \BibitemOpen
  \bibfield  {author} {\bibinfo {author} {\bibfnamefont {B.}~\bibnamefont
  {Julsgaard}}, \bibinfo {author} {\bibfnamefont {C.}~\bibnamefont {Grezes}},
  \bibinfo {author} {\bibfnamefont {P.}~\bibnamefont {Bertet}}, \ and\ \bibinfo
  {author} {\bibfnamefont {K.}~\bibnamefont {M{\o}lmer}},\ }\href {\doibase
  10.1103/PhysRevLett.110.250503} {\bibfield  {journal} {\bibinfo  {journal}
  {Phys. Rev. Lett.}\ }\textbf {\bibinfo {volume} {110}},\ \bibinfo {pages}
  {250503} (\bibinfo {year} {2013})}\BibitemShut {NoStop}%
\bibitem [{\citenamefont {Simon}\ \emph {et~al.}(1987)\citenamefont {Simon},
  \citenamefont {Sudarshan},\ and\ \citenamefont
  {Mukunda}}]{Simon.PhysRevA.36.3868(1987)}%
  \BibitemOpen
  \bibfield  {author} {\bibinfo {author} {\bibfnamefont {R.}~\bibnamefont
  {Simon}}, \bibinfo {author} {\bibfnamefont {E.~C.~G.}\ \bibnamefont
  {Sudarshan}}, \ and\ \bibinfo {author} {\bibfnamefont {N.}~\bibnamefont
  {Mukunda}},\ }\href {\doibase 10.1103/PhysRevA.36.3868} {\bibfield  {journal}
  {\bibinfo  {journal} {Phys. Rev. A}\ }\textbf {\bibinfo {volume} {36}},\
  \bibinfo {pages} {3868} (\bibinfo {year} {1987})}\BibitemShut {NoStop}%
\bibitem [{\citenamefont {Lobino}\ \emph {et~al.}(2008)\citenamefont {Lobino},
  \citenamefont {Korystov}, \citenamefont {Kupchak}, \citenamefont {Figueroa},
  \citenamefont {Sanders},\ and\ \citenamefont
  {Lvovsky}}]{Lobino.Science.322.563(2008)}%
  \BibitemOpen
  \bibfield  {author} {\bibinfo {author} {\bibfnamefont {M.}~\bibnamefont
  {Lobino}}, \bibinfo {author} {\bibfnamefont {D.}~\bibnamefont {Korystov}},
  \bibinfo {author} {\bibfnamefont {C.}~\bibnamefont {Kupchak}}, \bibinfo
  {author} {\bibfnamefont {E.}~\bibnamefont {Figueroa}}, \bibinfo {author}
  {\bibfnamefont {B.~C.}\ \bibnamefont {Sanders}}, \ and\ \bibinfo {author}
  {\bibfnamefont {A.~I.}\ \bibnamefont {Lvovsky}},\ }\href {\doibase
  10.1126/science.1162086} {\bibfield  {journal} {\bibinfo  {journal}
  {Science}\ }\textbf {\bibinfo {volume} {322}},\ \bibinfo {pages} {563}
  (\bibinfo {year} {2008})}\BibitemShut {NoStop}%
\bibitem [{\citenamefont {Kim}\ \emph {et~al.}(1989)\citenamefont {Kim},
  \citenamefont {\protect{de Oliveira}},\ and\ \citenamefont
  {Knight}}]{Kim.PhysRevA.40.2494(1989)}%
  \BibitemOpen
  \bibfield  {author} {\bibinfo {author} {\bibfnamefont {M.~S.}\ \bibnamefont
  {Kim}}, \bibinfo {author} {\bibfnamefont {F.~A.~M.}\ \bibnamefont
  {\protect{de Oliveira}}}, \ and\ \bibinfo {author} {\bibfnamefont {P.~L.}\
  \bibnamefont {Knight}},\ }\href {\doibase 10.1103/PhysRevA.40.2494}
  {\bibfield  {journal} {\bibinfo  {journal} {Phys. Rev. A}\ }\textbf {\bibinfo
  {volume} {40}},\ \bibinfo {pages} {2494} (\bibinfo {year}
  {1989})}\BibitemShut {NoStop}%
\bibitem [{\citenamefont {Gong}\ and\ \citenamefont
  {Aravind}(1990)}]{Gong.AmJPhys.58.1003(1990)}%
  \BibitemOpen
  \bibfield  {author} {\bibinfo {author} {\bibfnamefont {J.~J.}\ \bibnamefont
  {Gong}}\ and\ \bibinfo {author} {\bibfnamefont {P.~K.}\ \bibnamefont
  {Aravind}},\ }\href {\doibase 10.1119/1.16337} {\bibfield  {journal}
  {\bibinfo  {journal} {Am. J. Phys.}\ }\textbf {\bibinfo {volume} {58}},\
  \bibinfo {pages} {1003} (\bibinfo {year} {1990})}\BibitemShut {NoStop}%
\bibitem [{\citenamefont {Gardiner}\ and\ \citenamefont
  {Collett}(1985)}]{Gardiner.PhysRevA.31.3761(1985)}%
  \BibitemOpen
  \bibfield  {author} {\bibinfo {author} {\bibfnamefont {C.~W.}\ \bibnamefont
  {Gardiner}}\ and\ \bibinfo {author} {\bibfnamefont {M.~J.}\ \bibnamefont
  {Collett}},\ }\href {\doibase 10.1103/PhysRevA.31.3761} {\bibfield  {journal}
  {\bibinfo  {journal} {Phys. Rev. A}\ }\textbf {\bibinfo {volume} {31}},\
  \bibinfo {pages} {3761} (\bibinfo {year} {1985})}\BibitemShut {NoStop}%
\bibitem [{\citenamefont {Bouwmeester}\ \emph {et~al.}(2000)\citenamefont
  {Bouwmeester}, \citenamefont {Pan}, \citenamefont {Weinfurter},\ and\
  \citenamefont {Zeilinger}}]{Bouwmeester.JModOpt.47.279(2000)}%
  \BibitemOpen
  \bibfield  {author} {\bibinfo {author} {\bibfnamefont {D.}~\bibnamefont
  {Bouwmeester}}, \bibinfo {author} {\bibfnamefont {J.-W.}\ \bibnamefont
  {Pan}}, \bibinfo {author} {\bibfnamefont {H.}~\bibnamefont {Weinfurter}}, \
  and\ \bibinfo {author} {\bibfnamefont {A.}~\bibnamefont {Zeilinger}},\ }\href
  {\doibase 10.1080/09500340008244042} {\bibfield  {journal} {\bibinfo
  {journal} {J. Mod. Opt.}\ }\textbf {\bibinfo {volume} {47}},\ \bibinfo
  {pages} {279} (\bibinfo {year} {2000})}\BibitemShut {NoStop}%
\bibitem [{\citenamefont {Gardiner}\ and\ \citenamefont
  {Zoller}(2000)}]{Gardiner.QuantumNoise}%
  \BibitemOpen
  \bibfield  {author} {\bibinfo {author} {\bibfnamefont {C.~W.}\ \bibnamefont
  {Gardiner}}\ and\ \bibinfo {author} {\bibfnamefont {P.}~\bibnamefont
  {Zoller}},\ }\href@noop {} {\emph {\bibinfo {title} {Quantum Noise}}},\
  \bibinfo {edition} {2nd}\ ed.\ (\bibinfo  {publisher} {Springer},\ \bibinfo
  {address} {Berlin},\ \bibinfo {year} {2000})\BibitemShut {NoStop}%
\bibitem [{\citenamefont {Julsgaard}\ and\ \citenamefont
  {M{\o}lmer}()}]{Julsgaard.arXiv.1309.5517(2013)}%
  \BibitemOpen
  \bibfield  {author} {\bibinfo {author} {\bibfnamefont {B.}~\bibnamefont
  {Julsgaard}}\ and\ \bibinfo {author} {\bibfnamefont {K.}~\bibnamefont
  {M{\o}lmer}},\ }\href@noop {} {}\bibinfo {howpublished}
  {arXiv:1309.5517}\BibitemShut {NoStop}%
\end{thebibliography}

% Paste from article.bbl:
%

\end{document}